%% file: ms.tex
\documentclass[12pt,preprint]{aastex}
\shorttitle{Cold dust in late-type Virgo Cluster galaxies}
\shortauthors{V\"olk et al. }
\usepackage{graphicx}
\usepackage{subfigure}

\include{eps}

\begin{document}

\title{Cold dust in late-type Virgo Cluster galaxies} 

\author{Cristina C. Popescu\altaffilmark{1,2}}
\affil{The Observatories of the Carnegie Institution of Washington, 
813 Santa Barbara Str., Pasadena, 91101 California, USA}
\author{Richard J. Tuffs, Heinrich J. V\"olk}
\affil{Max Planck Institut f\"ur Kernphysik, Saupfercheckweg 1, 
69117 Heidelberg, Germany} 
\author{Daniele Pierini}
\affil{The University of Toledo, Toledo, Ohio 43606-3390, USA}
\author{Barry F. Madore\altaffilmark{3}} 
\affil{NASA/IPAC Extragalactic Database, 770 S. Wilson Avenue, Pasadena, 
California 91125, USA}

\altaffiltext{1}{Max Planck Institut f\"ur Astronomie, K\"onigstuhl 17, 69117
Heidelberg, Germany}
\altaffiltext{2}{Research Associate, The Astronomical Institute of the 
Romanian Academy, Str. Cu\c titul de Argint 5, Bucharest, Romania}
\altaffiltext{3}{The Observatories of the Carnegie Institution of Washington, 
813 Santa Barbara Str., Pasadena, 91101 California, USA}

\begin{abstract}

We have statistically analyzed the spatially integrated 
far-infrared (FIR) emissions of the complete volume- and 
luminosity-limited sample of late-type (later than S0) Virgo cluster 
galaxies  measured using the Infrared Space Observatory (ISO)
by Tuffs et al. (2002) in bands centered on 60, 100 \& 170\,${\mu}$m.
30 out of 38 galaxies detected at all three wavelengths contain 
a cold dust emission component, present within all morphological 
types of late-type systems ranging from early giant spirals to 
Blue Compact Dwarfs (BCDs), and which could not have been recognized 
by IRAS.
We fitted the data with a superposition of two modified grey-body 
functions, physically identified 
with a localized warm dust emission component associated with HII 
regions (whose temperature was constrained to be 47\,K), and a 
diffuse emission component of cold dust. 
The cold dust temperatures
were found to be broadly distributed, with a median of 18\,K, some 
$8-10$\,K lower than would have been predicted from IRAS. 
The derived total dust masses are correspondingly increased by factors 
of typically $6 - 13$. 
A good linear correlation is found between the ``warm FIR'' 
luminosities and the H${\alpha}$ equivalent widths (EW), supporting 
the assumptions of our constrained spectral energy distribution (SED) 
fit procedure. We also found a 
good non-linear correlation between the ``cold FIR'' luminosities
and the H${\alpha}$ EWs, consistent with the prediction of Popescu
et al. (2000) that the FIR-submm emission should mainly be due to
diffuse non-ionizing UV photons. Both the ``warm'' and the ``cold''
FIR luminosity components are non-linearly correlated with the 
(predominantly non-thermal) radio luminosities.  
There is a tendency for the temperatures of the cold dust component
to become colder, and for the cold dust surface densities 
(normalized to optical area) to increase for later morphological 
types. A particularly significant result concerns the low dust temperatures 
(ranging down to less than $10$\,K) and large dust masses associated 
with the Im and BCD galaxies in our sample. We propose two scenarios
to account for the FIR characteristics of these systems.

\end{abstract}


\section{Introduction}

IRAS observations led to the belief that dust emission in 
normal galaxies comes from $\sim30$\,K dust grains with a total gas-to-dust 
mass ratio of $\sim10^3$ (e.g. Devereux \& Young 1990). This is about one 
order of magnitude greater than the gas-to-dust mass ratio our Galaxy, a 
discrepancy which suggested  that most of the dust in late-type galaxies had 
in fact been ``overlooked'' by IRAS. Indeed, one obvious bias of IRAS studies 
in general is the lack of 
spectral coverage longwards of the 100\,${\mu}$m filter and the three times 
brighter sensitivity limit in this band compared to the IRAS 60\,${\mu}$m 
band. This might translate into a bias against the detection of cold dust 
with temperatures colder than 30\,K.  

The first observations suggesting the existence  of a cold dust component
in galaxies were made at sub-millimeter (sub-mm) wavelengths by Chini et 
al. (1986) using the 3\,m IRTF (Infrared Telescope Facility) and the 
UH 88$^{\prime\prime}$ telescope (University of Hawaii). The strong sub-mm 
emission found
in all the observed galaxies could not be accounted for by dust emitting at
a uniform temperature and the FIR spectra were interpreted in terms of two
dust components of about 16 and 53\,K. This interpretation suggested that the
peak of the FIR emission should be at wavelengths between 100 and 
200\,${\mu}$m. Subsequent sub-mm
observations done with the 30\,m IRAM (Pico Veleta) telescope also suggested
the presence of a cold dust component in individual nearby galaxies: 
NGC~891 (Gu\'elin et al. 1993), NGC~3627 (Sievers et al. 1994), NGC~4631
(Braine et al. 1995), M~51 (Gu\'elin et al. 1995), 
NGC~4565 (Neininger et al. 1996), NGC~3079 (Braine et al. 1997), 
NGC~5907 (Dumke et al. 1997). The enhanced sensitivity and overall efficiency
improvement of the SCUBA sub-mm array at the JCMT 
(James Clerk Maxwell Telescope) extended our knowledge of the quantity of cold
dust by probing both more ``quiescent'' galaxies and also by mapping its
distribution with higher resolution (Alton et al. 1998a, Israel et al. 1999, 
Bianchi et al. 2000b, Alton et al. 2001). However, none of these observations 
directly observed the spectral peak of the FIR emission and thus 
did not unambiguously distinguish between the two dust components. 
Observations at 160 and 200\,${\mu}$m were until 
recently available only for a few bright objects and were performed with the 
KAO (NASA-Kuiper Airborne Observatory)(e.g. Engargiola 1991). 

The ISOPHOT instrument (Lemke et al. 1996) on board the ISO satellite 
(Kessler et al. 1996) extended for the first time the
wavelength coverage beyond that of IRAS, to 240\,${\mu}$m, with superior 
intrinsic sensitivity, and the availability of longer integration times than 
were possible with IRAS. ISO was thus capable to cover the peak in energy 
distribution ${\nu}$S$_{\nu}$ for normal galaxies, detecting discrete sources 
at least 10 times fainter than IRAS at 60 and 100\,${\mu}$m. First results 
from ISOPHOT have indeed confirmed the existence of a cold dust component in 
several nearby normal galaxies: NGC~6946 (Tuffs et al. 1996), M~31 
(Haas et al. 1998), or on small samples: Kr\"ugel et al. (1998), 
Siebenmorgen, Kr\"ugel \& Chini (1999),Contursi et al. (2001). First results 
from the ISO LWS (Long Wavelength Spectrograph) (Trewhella et al. 2000) are 
also consistent with the existence of cold dust in normal galaxies. 
The sample of compact sources with galaxy associations
from the ISOPHOT 170\,${\mu}$m serendipity survey (Stickel et al. 2000)
presents statistical evidence for a cold dust component, but the sample
was obviously biased towards FIR luminous systems, with uncertain transient 
corrections in the observations. 

In this paper we present statistically significant evidence for the 
existence of a cold dust component in all galaxies later than S0, i.e. 
spirals, irregulars and Blue Compact Dwarfs (BCDs). This result comes from our 
analysis of a complete volume- and luminosity-limited sample of late-type 
Virgo Cluster galaxies observed by Tuffs et al. (2002) with the ISOPHOT 
instrument. Our sample is intended to give a more representative statistical 
analysis of 
the cold dust component, for a full range in morphological type, and to 
reach fainter detection limits than previously available. In addition, we 
present our analysis on data which are for the first time corrected from the 
transient effects of the detectors, allowing a quantitative and comprehensive 
evaluation of the cold dust on a relatively deep and unbiased sample.

A main goal for our study is the detailed knowledge of the shape of the FIR 
SED in galaxies, as well as the distribution and morphology of the cold dust. 
This is required for understanding the energy budget in galaxies and for 
modeling the SED over the whole wavelength range. Dust grains can be 
considered as test particles for the intrinsic radiation fields in galaxies. 
Therefore observations of their emission in the FIR, combined with optical and
ultraviolet (UV) data of the light from stars, attenuated by the grains, 
should, in principle, strongly constrain the intrinsic distribution of 
stellar luminosity and dust 
in galaxies. This would address the fundamental question of the optical 
thickness of galactic disks and would allow evaluation of intrinsic quantities 
of interest - the star formation rate (SFR) and the star formation history. 
This is possible only by combining the new FIR observational results with 
modeling techniques for the SED from the UV to the FIR/sub-mm range. 
Different tools have been
proposed for analyzing the energy budget of normal galaxies, starting with the
pioneering works of Xu \& Buat (1995) and Xu \& Helou (1996) and continuing 
with more recent models, like those of Silva et al. (1998), 
Devriendt et al. (1999), Bianchi et al. (2000a). However, most of these models 
have made severe simplifications, which ultimately led to contradictory 
conclusions on the disk opacities and on the origin of the FIR emission. 

Recently a new tool for the analysis of the SED in galaxies has been proposed 
by Popescu et al. (2000). This tool includes solving the radiative-transfer 
problem for a realistic distribution of absorbers and emitters, 
considering realistic models for dust (taking into account the grain size 
distribution and the stochastic heating of small grains) and the 
contribution of HII regions within star forming complexes. This tool was 
applied to the edge-on spiral galaxy NGC~891 with the result that most of
the cold dust that emits in the sub-mm was embedded in a second
disk of dust, associated with the young stellar population and having a very
small scale height ($\sim 90$\,pc). Subsequent application of this tool to a
sample of 5 edge-on spiral galaxies (Misiriotis et al. 2001) has
shown that the model seems to be generally valid and that the 
interpretation of the cold dust component should be ultimately analyzed in 
terms of SED-modeling techniques.

Another astrophysical problem related to cold dust in galaxies is the source 
of heating of this dust component. Here again, modeling
techniques are needed for a detailed and better understanding of this 
component. Xu \& Buat (1995) claimed that the non-ionizing UV photons 
constitute the main heating source of dust. However, their results were applied
only to the IRAS bands, which did not reveal the bulk of the dust mass in
normal galaxies, which, as we show in this paper, is in most cases
too cold to have been seen by IRAS. With the advantage of the longer 
wavelengths observations, Popescu et al. (2000) have shown that, for the case 
of NGC~891, the 
dust emitting in the $170-850$\,${\mu}$m
regime is also predominantly heated by the UV radiation from the young stellar 
population.
More fundamental is the accompanying result showing that the FIR colors have 
to be interpreted also in
geometrical terms, rather than simply as separate temperature components.

The morphology of the cold dust component and its geometrical
distribution is a primary goal of the ISOPHOT Virgo project. 
Especially of interest is the
question of whether the cold dust component extends beyond the optical disk 
of the galaxies, as predicted by the radiative transfer modeling analysis of 
the optical data of Xilouris et al. (1997, 1998, 1999) or as suggested by 
Alton et al. (1998a) from the analysis of the sub-mm data. 
Similar suggestions (Davies et al. 1999, Alton et al. 1998b) have been made 
on the basis of ISOPHOT maps of several nearby galaxies. However, these maps 
were not corrected for transient effects of the detector, and therefore both 
the calibration and the derived scale lengths are uncertain, and they are 
subject to instrumental effects.  

The paper is organized as follows. Sect.~2 presents the main
characteristics of our sample used for the statistical analysis of
the cold dust component. The observed colors are presented in Sect.~3.
In Sect.~4 dust temperatures, masses and luminosities are derived from 
fitting the SED of our sample galaxies. The distributions in the main derived 
quantities are presented in Sect.~5, and their dependence on Hubble type is 
investigated in Sect.~6. Sect.~7 presents the correlation of the FIR 
luminosity with indicators of the SFR. The implications of our analysis of 
the cold dust content of Virgo galaxies for our understanding of normal 
galaxies is discussed in Sect.~8. The summary and conclusions are given in
Sect.~9. 

\section{Sample definition and photometric accuracy}

Our sample comprises the late-type Virgo Cluster galaxies observed with the
ISOPHOT instrument on board the ISO satellite. Fuller details of the 
selection criteria, observations, data reductions, and the catalog itself 
with the resulting photometry is presented in an accompanying paper 
(Tuffs et al. 2002). Here we give only a short description of the main 
aspects of the sample definition and FIR photometry relevant for the
statistical investigation of cold dust in galaxies.

The observed sample consists of 63 member galaxies selected from the Virgo 
Cluster catalogue (VCC) (Binggeli, Sandage \& Tammann 1985), with Hubble 
type later than S0. The galaxies were chosen to maximize the likelyhood of 
their belonging to the main cluster, thus minimizing the spread in distance 
due to the complex 3D structure (Binggeli, Popescu \& Tammann 1993 and
references therein). The sample is divided into cluster-core and 
cluster-periphery subsamples, complete to $B_{\rm T} = 14.5$ and $16.5$, 
respectively. The 35 observed periphery galaxies are probably freshly 
falling in from the field (see Tully \& Shaya 1984). They are principally 
comprised of spirals later than Sbc, irregulars and BCDs. The 28 observed 
core galaxies are essentially seen towards the extended X-ray halo of M~87 
and are dominated by spirals of type Sc and earlier.

The observations were done using the C100 and C200 detectors, in ISOPHOT's  
``P32'' observing mode (Tuffs \& Gabriel 2001), which uses the focal plane 
chopper in conjunction with a spacecraft raster to rapidly sample
large areas of sky. The observing wavelengths were 60, 100 and 
170\,${\mu}$m. The ``P32'' mode allowed
the entire optical extent of each target down to the
25.5\,mag\,arcsec$^{-2}$ B-band isophote and adjacent
background to be scanned, while still maintaining a
spatial oversampling. This allowed both spatially integrated
far-IR photometry as well as information on the morphology of the
galaxies in the far-IR to be extracted from the data.
This paper is primarily concerned with the integrated photometry, which
will include any emission from cold dust in the outer disks, where
the radiation fields are weak. The most important aspect of the data 
reduction described by Tuffs et al. (2002) was the correction of the complex
transient response behavior of the Ge:Ga photoconductor detectors
of ISOPHOT. Failure to correct for this effect in data taken in
the ``P32'' mode can give rise to serious signal losses and distortions
in the derived brightness profiles through the galaxies, which in
turn can lead to spurious results on the amount and
distribution of cold dust in these systems. Our data
are the first from the ``P32'' mode to have been corrected for these
effects. As demonstrated by Tuffs et al. (2002) the transient-corrected
photometry correlates well with IRAS observations of the brighter
galaxies in our sample in the ISO 60 and 100\,${\mu}$m pass bands
over about two orders of magnitude in integrated flux density.

From the 63 galaxies observed
(61 galaxies at all three FIR wavelengths and 2 galaxies 
only at 100 and 170\,${\mu}$m) we detected 54 galaxies at least at one 
wavelength and 40 galaxies at all three wavelengths. From the 40 galaxies
detected at all FIR wavelengths, 2 galaxies form an interacting pair 
(VCC~1673/1676), and their properties will be discussed in a separate 
paper. In this paper we largely restrict our analysis to the remaining 38 
galaxies with detections at 60, 100 and 170\,${\mu}$m. One galaxy, 
VCC~1110,  was observed at two additional wavelengths, 70 and 120\,${\mu}$m, 
in order to have a more detailed knowledge of the shape of the SED. This 
sample forms the basis for our statistical investigation. In Sect.~6, where 
we discuss the variation of the FIR properties with Hubble type, we add to 
our sample of 38 galaxies the 3 galaxies having detections only at 100 and 
170\,${\mu}$m. The latter are introduced to increase the statistics of the 
BCDs (since all 3 extra galaxies are BCDs). The flux densities and their associated errors are taken 
from Table~7 of Tuffs et al. (2002), while the morphological types and other 
optical properties of the sample galaxies can be found in Table~1 of the 
same paper.

\section{Flux density ratios}

Fig.~1a shows the color-color diagram log(F170/F100) vs log(F100/F60), with the
60 and 100\,${\mu}$m ISO flux densities converted to the IRAS 
scale.\footnote{As shown by Tuffs et al. (2002), the flux scale of the ISOPHOT 
Virgo survey has a relative gain ISO/IRAS of 0.95 and 0.82 at 60 and 
100\,${\mu}$m, respectively. These corrections have been applied to the flux 
densities plotted in this figure only to facilitate comparison with IRAS 
colors.}
There is no obvious correlation, but rather a scatter diagram with the 
logarithm of F100/F60 ranging between $\sim$0 and 0.9 and with the logarithm 
of F170/F100 ranging between $\sim$0 and 0.6. Three galaxies have unusual 
warmer F170/F100 colors, with negative logarithmic ratios, below 0. The large 
scatter in the F100/F60 color can be
interpreted as evidence for of a large range in star formation activity. The
large scatter in the F170/F100 color indicates that cold dust is present with a
large variation in dust temperatures. There is no obvious
segregation with morphological type, though
the galaxy with the coldest F170/F100 color is a BCD and the galaxies with the
warmest F170/F100 colors are early type spirals. 

Fig.~1b shows for comparison the color-color diagram of a sample of 
compact sources with galaxy associations detected in the ISOPHOT 170\,${\mu}$m 
serendipity survey (Stickel et al. 2000). This is the only statistical 
sample of galaxies observed at 170\,${\mu}$m existing in the literature
which is also comprised of a large variety of morphological types. We will 
refer to this sample as to the ``serendipity sample''. Nevertheless the 
selection criteria of our sample is quite different from
that of the ``serendipity sample''. First of all our sample is a cluster 
sample, while the ``serendipity sample'' is mainly representative of the 
field population. Our sample is a volume- and luminosity-limited sample, 
selected from an optical catalogue, while the ``serendipity sample'' is a 
blind survey at 170\,${\mu}$m, which will predominantly be biased towards 
luminous FIR sources. The redshift distribution of the latter sample 
(see their Fig.~4) shows that, although the majority of sources have low 
redshifts of $z<0.02$, there is a long tail of redshifts up  
to $z\approx0.05$. Despite the differences in the selection criteria used for
the two samples, it is still useful to compare the samples and investigate
their FIR properties. 

The color-color diagram of the ``serendipity sample' shows a slight tendency 
for warmer F100/F60 colors than the Virgo sample and a somewhat smaller 
(though still substantial) scatter in both colors. 
The larger scatter in our sample 
cannot be explained as being due to systematic calibration
uncertainties, nor due to random errors (see Fig.~1a).
This appears to be a real effect and one could speculate that, if there is a 
FIR color - FIR luminosity relation, or a relation between spread in FIR 
color over a population and FIR luminosity, then a sample biased towards 
higher FIR luminosities (as we expect the ``serendipity sample'' to be) 
will not reflect the full intrinsic spread in colors that a deeper sample 
embracing a larger proportion of low luminosity objects would. This would be
consistent with more active star-forming galaxies to have narrower, warmer
FIR SEDs.

\section{The spectral energy distribution}

The flux densities at 60, 100 and 170\,${\mu}$m (for a representative selection
see Fig.~2\footnote{The SEDs of all galaxies from our sample can be found at
\newline$http://nedwww.ipac.caltech.edu/level5/Sept01/Popescu/Popescu\_contents.html$}) 
indicate that for most of the galaxies of our sample the SED in 
the FIR peaks at wavelengths $>100\,{\mu}$m and cannot be represented by a 
single modified 
black-body (Planck) function. The turn-up in the SED beyond 100\,${\mu}$m 
is a clear indication of a cold dust component which could not have been 
detected by IRAS. Perhaps the simplest way of quantifying this cold 
component would be to fit the FIR SED with 2 modified blackbody functions, 
one representing the so-called ``warm dust'' component and the other 
representing the so-called ``cold dust'' component. However, such a fit 
requires 4 free parameters (the amplitudes and temperatures of the ``warm'' 
and ``cold'' components, respectively) while there are only 3 data points
available. Obviously there is an infinity of solutions which would fit the
data, since the problem is mathematically under-constrained. Sub-mm data would
alleviate this problem. However, long wavelength observations are not yet
available for our Virgo sample. For one galaxy, VCC~1110, observations
with the ISOPHOT C70 and C120 filters were made to obtain a more detailed
knowledge on the shape of the SED. For this particular galaxy the fit to the
data can provide a unique solution; for all the other galaxies
the model fit is not unique. One way to deal with this problem is to fit a
modified black-body curve only to the 100 and 170\,${\mu}$m data, and obtain a
lower limit for the amount of cold dust and an upper limit for its 
temperature. Another possibility is to try to constrain the problem
using some physical considerations. We use both procedures and show that they
lead to the same statistical results, the differences being in the zero point
of the estimated dust masses. 

We first fitted the 100 and 170\,${\mu}$m data with one modified black-body
function:
\begin{eqnarray}
F_{\nu} \sim {\nu}^{\beta} B_{\nu} (T_{\rm D})
\end{eqnarray}
with a fixed emissivity index ${\beta}=2$. Since the ISOPHOT flux densities
refer to a spectrum with ${\nu}F_{\nu}=$\,constant, color corrections were 
first applied to the data. From the amplitude of the fitted modified 
black-body function dust masses were derived using the formula:
\begin{eqnarray}
M_{\rm d} = \frac{4}{3}\,\frac{F_{\nu}\,\rho\,d^2}{Q_{\rm i}}
\lambda^{2}\,B_{\nu}^{-1}(T_{\rm D})
\end{eqnarray}
where $\rho$ is the density of the grain material 
($3.2\,{\rm g}\,{\rm cm}^{-3}$ for silicates and 
$2.3\,{\rm g}\,{\rm cm}^{-3}$ for graphites), $d$ is the distance to the
galaxy, and $Q_{\rm i}$ is the emissivity constant (130\,${\mu}$m for 
silicates and 230\,${\mu}$m for graphites, derived for the FIR/submm regime 
from Draine 1985 and consistent with the adopted emissivity index $\beta=2$). Dust masses were calculated under the assumption of an 
interstellar mixture of silicates and graphites. The graphite and silicate
abundances were taken from Draine \& Lee (1984), namely 53$\%$ silicates 
($N_{\rm Si}$) and 47$\%$ graphites ($N_{\rm Graphite}$) which were chosen to fit the
extinction curve in our Galaxy and which we also adopted here. Dust masses and
temperatures were derived for all our sample galaxies with detected 100 and
170\,${\mu}$m flux densities which show evidence for two dust temperature 
components.  In deriving dust masses and luminosities we assume that all the 
galaxies have the same distance, namely $11.5\times {\rm h}^{-1}$\,Mpc 
(see Binggeli, Popescu \& Tammann 1993),  where ${\rm h}={\rm H_0}/100$ and 
H$_0$ is the Hubble constant. 


\begin{table}
\tablenum{1}
\caption{Examples of the dependence of $T_{\rm D}^{\rm cold}$ on 
$T_{\rm D}^{\rm warm}$}

\begin{tabular}{r|rrr}
\hline
VCC & $T_{\rm D}^{\rm warm}=46.0$\,K & $T_{\rm D}^{\rm warm}=47.0$\,K & 
$T_{\rm D}^{\rm warm}=48.0$\,K \\
\hline
152 & $T_{\rm D}^{\rm cold}=18.5$\,K & $T_{\rm D}^{\rm cold}=18.5$\,K & 
$T_{\rm D}^{\rm cold}=18.4$\,K \\
    & $\sigma=0.5$\,K & $\sigma=0.5\,$K & $\sigma=0.5$\,K \\
318 & $T_{\rm D}^{\rm cold}=14.9$\,K & $T_{\rm D}^{\rm cold}=14.9$\,K & 
$T_{\rm D}^{\rm cold}=14.9$\,K \\
    & $\sigma=0.5$\,K & $\sigma=0.5$\,K & $\sigma=0.5$\,K \\
\hline
\end{tabular}
\end{table}

Secondly we fitted the FIR SED with two modified black-body functions by
constraining one parameter of the fit. The most likely constraint is to fix the
temperature of the ``warm dust'' component.
As shown by Popescu et al. (2000) for the
nearby spiral NGC~891, most of the emission at 60\,${\mu}$m is due to localized
FIR  sources within HII regions and star forming complexes ($62\%$). Since 
these results seem to be valid also for other disk galaxies
(Misiriotis et al. 2001) we fixed the temperature of the ``warm dust''
component by making the assumption that this will represent the
temperature of the average HII regions within each galaxy. Obviously this
assumption neglects the fact that part of the FIR emission at 60\,${\mu}$m is
also due to dust heated by the diffuse optical radiation in the center of the
disk (20\% for NGC~891, see Popescu et al. (2000)) and by the diffuse UV
radiation field in the outer parts
of the disk, where small grains are stochastically heated ($19\%$ for
NGC~891, see Popescu et al. (2000)). 
The temperature of
the warm component was fixed to 47\,K, as it provides the best fits and minimal
uncertainty in the fitted parameters for all galaxies. In practice, even 
relatively large deviations from the 47\,K temperature of the warm component 
will not introduce large deviations in the fitted temperature of the cold 
component; that is, the cold dust component is quite stable against the 
particular shape of the warm dust component. To demonstrate this we changed the
temperature of the warm component to 46 and 48\,K, respectively. Table~1 
 shows the results for two galaxies, VCC~152 and VCC~318. The
resulting fitted temperatures of the cold component and associated 
uncertainties are similar. Finally, the quantities derived from this
procedure are compared with the fitted parameters from the one modified 
black-body fit.

The flux densities were again color-corrected and dust masses and temperatures
were derived for all sample galaxies with detections at all three wavelengths
using the same procedure outlined for one modified black-body fit. The 
results of the fits are listed in Table~2 and some representative
examples of fitted FIR SEDs are plotted in Fig.~2. For 
the case of VCC~1110, where observations in more than 3 filters were available,
 we have used the same constrained fit and compared the predictions of the 
model fit with the available data at 70 and 120\,${\mu}$m, not used in the fit.
Fig.~2 shows a very good agreement between the model
predictions and the observations for this galaxy. This test is reassuring for 
the use of our procedure.


In order to establish whether different fitting procedures introduce internal
scatter in the derived parameters of the fit we cross-correlated the dust
masses obtained from fitting a single modified black-body function to the 100 
and 170\,${\mu}$m flux densities with the dust masses of the cold component
obtained from the two modified black-body fit. 
The correlation was tested for the galaxies
with detections at all three wavelengths and which presented evidence for two
dust temperature components. Fig.~3 shows a very
tight correlation between the two dust masses, with only a small scatter for
low dust masses (correlation coefficient 0.987). Obviously the correlation 
shows that using a constrained fit
will give the same statistical results as the lower limits estimates, with the
exception of an uncertainty in the zero point of the dust masses. Therefore for
the rest of the paper we will use only the results from the constrained fit.

For 8 galaxies in our sample there was no evidence for two dust
components with different temperatures, and their SED was fitted with a 
single modified 
black-body function. The parameters of these fits are listed in Table~3. 
These peculiar galaxies 
have either a very warm SED which peaks around 100\,${\mu}$m 
(VCC~836/1003/1326) or have very cold F100/F60 colors 
(VCC~1253/1419/1450/1725/1757)  such that the 60\,${\mu}$m data 
would lie on the same modified black-body curve which is defined by the 100 
and 170\,${\mu}$m flux densities. Most of these one temperature dust component
galaxies are early-type spirals in the cluster core, some of them with Sy 2
activity (VCC836/1253) or with peculiar morphologies 
(VCC~1419/1757). Throughout this paper we will refer to these galaxies as
to the ``one component'' galaxies.

\section{The distributions of dust temperatures, masses and luminosities}

\subsection{Dust temperatures}

The distribution of the cold dust temperatures ($T^{\rm cold}_{\rm D}$) is 
shown in Fig.~4a. The ``one component''
galaxies have a distinct locus in the histogram, since they form the highest
temperature tail of the distribution, with values between 20 and 28\,K. The
most extreme example is VCC~1326 (a SBa galaxy in the cluster core) with
 $T_{\rm D}=34.7$\,K. For the ``two component'' galaxies the temperature 
distribution shows a wide spread in values, 
ranging from 12\,K to 21\,K. The coolest object from the distribution 
is VCC~655 with $T_{\rm D}^{\rm cold}=12.6$\,K. This galaxy is classified as
a S pec/BCD in the cluster periphery. 

The dust temperatures derived for our sample are much lower than those IRAS
would have predicted on the basis of the 60 and 100\,${\mu}$m flux densities
only. If we consider our ISO flux densities at 60 and 100\,${\mu}$m
and only one component dust temperature we would derive a temperature
distribution with a median of 26\,K. This should be compared with the median 
value calculated for our cold dust temperature distribution, of 18.2\,K - if
the ``one component'' galaxies are included, or 16.7\,K - if these galaxies 
are excluded. Thus the cold dust component has a 
median value $8-10$\,K colder than IRAS would have predicted for our Virgo 
sample. 

\subsection{Dust masses}

Most of the dust content in our sample galaxies is in form of cold
dust. This can be seen from Table~2, where the tabulated cold dust masses
$M_{\rm D}^{\rm cold}$ are $3-4$ orders of magnitude larger than the
tabulated warm dust masses $M_{\rm D}^{\rm warm}$.  The distribution of cold
dust masses is shown in Fig.~4b. As in the temperature distribution, 
the ``one component'' galaxies have a distinct
locus in the histogram, occupying the lowest mass tail of the 
distribution, between 
$4.8 < \log M_{\rm D}^{\rm cold} < 6.3\,{\rm M}_{\odot}\times{\rm h}^{-2}$. 
For the remaining galaxies the distribution is again very broad, extending 
3 orders of magnitude in range, with a peak around 
$5.8\,{\rm M}_{\odot}\times{\rm h}^{-2}$. 
The galaxy with the largest amount of
cold dust ($1.14\times 10^8\,{\rm M}_{\odot}\times{\rm h}^{-2}$) is VCC~92, an
Sb galaxy in the cluster periphery which is also one of the biggest galaxies 
in our sample, perhaps indicative of a general scaling relation between total
dust mass and galaxy size. 

The dust masses derived for our sample are larger than those IRAS would have
predicted on the basis of the 60 and 100\,${\mu}$m flux densities alone. 
Following the same procedure as for the dust temperatures we derive an
``IRAS'' median value of $2.3\times 10^5\,{\rm M}_{\odot}\times {\rm h}^{-2}$ 
and an ISO median value of 
$1.3\times 10^6\,{\rm M}_{\odot}\times {\rm h}^{-2}$ - if the 
``one component'' galaxies are included, or 
$3.0\times 10^6\,{\rm M}_{\odot}\times {\rm h}^{-2}$ - if 
these galaxies are excluded. Thus our galaxy sample contains a
factor $6 - 13$ more dust (for the median value) than IRAS data alone would 
suggest. For individual galaxies this factor could be larger.

To compensate for the effect of scaling on the dust mass distribution we have
normalized dust masses in the form of dust mass surface 
densities,
$M^{\rm cold}_{\rm D}/D^2$, where D is the major diameter\footnote{The 
diameters were taken from the VCC catalogue (Binggeli, Sandage \& Tammann 
1985).} (in arcmin) of the galaxies measured to the faintest detectable 
optical surface brightness level of approximately 
$25.5\,{\rm B}\,{\rm mag}\,{\rm arcsec}^{-2}$. The
histogram of the dust mass surface densities (Fig.~4c) still shows a 
broad distribution, but not as broad as that in Fig.~4b. Except for the last 
histogram bin (one galaxy), the distribution of ``two component'' galaxies 
ranges over 2 orders of magnitudes, indicating an intrinsic variation in the
amount of cold dust within galaxies. Again, the ``one component'' galaxies
exhibit the lowest dust mass surface densities. The
galaxy with the largest content of cold dust with respect to its optical size 
 is VCC~655.

Another way of normalizing the dust masses is to take the ratio to the HI gas 
mass.\footnote{Most of the HI data used for
normalization were taken from Bottinelli et al. (1990) and 
Hoffman et al. (1987). For 3 galaxies (VCC~1043/1686/1690) the 
average of the measurements (beam corrected) existing in the literature was 
considered: Bottinelli et al. (1990), Guiderdoni \& Rocca-Volmerange (1985), 
Huchtmeier \& Richter (1986) for VCC~1043; Bottinelli et al. (1990), 
Hoffman et al. (1987), Guiderdoni \& Rocca-Volmerange (1985), 
Huchtmeier \& Richter (1986) for VCC~1686; Bottinelli et al. (1990), 
Guiderdoni \& Rocca-Volmerange (1985), Huchtmeier \& Richter (1986), 
Warmels 1988 for VCC~1690.
For VCC~1003, VCC~1253 and VCC~1326 the H I fluxes (beam corrected)
are upper limits and come from Huchtmeier \& Richter (1986).} As for the case
of dust mass surface densities, the distribution of 
dust-to-HI mass ratio $M_{\rm D}/M_{\rm HI}$ (Fig.~4d) 
for ``two component'' galaxies shows a broad distribution, ranging over two 
orders of magnitude. Unlike the dust mass surface densities, the 
distribution of dust-to-HI mass ratio for ``one component'' galaxies is 
more widely spread. The galaxy with the largest dust-to-HI mass ratio is 
VCC~655.

\subsection{Dust luminosities}

For most of our sample galaxies the FIR luminosity of 
the ``cold component'' is higher than the FIR luminosity of the 
``warm component'' by factors of between 1.3 and 4.1 (see Table~2). For two 
galaxies - VCC~655 and VCC~857 - the ratio between the cold and warm FIR 
luminosities is as high as 6.0 and 6.1, respectively. VCC~655 is the galaxy 
with the coldest temperature of the cold
dust component and the largest amount of dust with respect to its optical 
size. VCC~857 is a SBb galaxy in the cluster periphery with LINER activity. 
Despite its activity
the galaxy has the highest cold/warm FIR luminosity ratio. By contrast, 
there are two galaxies that radiate more FIR
luminosity in the warm component, namely VCC~1699 (a factor 1.2) and 
VCC~664 (a factor 2.1). VCC~1699 is a SBm galaxy in the cluster periphery and 
VCC~664 is a Sc galaxy in the cluster
core. One galaxy contributes equal FIR luminosity to the cold and warm 
component. This is the case for VCC~1554, an Sm galaxy in the cluster 
periphery.

The distribution of the integrated FIR luminosities for our sample galaxies 
(Fig.~5a) ranges over $34.0 < \log L_{\rm FIR} < 36.4\,{\rm W}\,{\rm h}^{-2}$.
The FIR luminosities of the ``cold component'' (Fig.~5b) seem to be more 
widely distributed than the FIR luminosities of the ``warm component'' 
(Fig.~5c). This could be partly a consequence of the constraint in the
fitting procedure, namely the adopted fixed temperature of the warm component. 
After normalization to the K$^\prime$ band magnitudes\footnote{The total 
K$^\prime$ band magnitudes used for normalization were derived from the 
observations of Boselli et al. (1997) and corrected for Galactic extinction 
and inclination, according to the K-band corrections, as described in 
Gavazzi \& Boselli (1996). Their median uncertainty is 0.15 mag.} the 
distribution of the FIR luminosities (Fig.~5d) becomes slightly narrower for 
the ``two component'' galaxies. There is also a hint that the distribution 
of cold dust normalized luminosity (Fig.~5e) is narrower than that of the 
warm dust normalized luminosity (Fig.~5f). Apart from a scaling effect this 
might be due to a contribution of the old stellar population to the heating 
of cold grains. 

\section{FIR properties with respect to Hubble type}

To study the FIR properties with respect to Hubble type
we divide our sample into 4 subsamples corresponding to S0a-Sa, Sab-Sc, Scd-Sm
and Im-BCDs bins in Hubble type. Ideally one would like to have very good
statistics within each bin in Hubble type. Our sample has obviously the best
statistics within the Sab-Sc bin and poor statistics for the
Im-BCD bin. The poor statistics in the latter bin is mainly a consequence of
the fact that some BCDs are only detected at two wavelengths, or only at
one wavelength, and were thus not included in the overall statistics. To 
improve the statistics of the BCDs and to highlight their unusual properties
we have added to our sample the galaxies detected only at two wavelengths 
(100 and 170\,${\mu}$m), namely VCC~130/848/1750, all of BCD
class. Dust temperatures derived for these galaxies are upper limits and dust 
masses are lower limits (see Table~3).

The temperatures of the cold dust component
(Fig.~6a) have a tendency to become colder for the later types. The
early-type spirals have a temperature distribution shifted towards
higher dust temperatures, with the coldest temperature only 17.7\,K and
the warmest 33.4\,K. The broadest temperature distribution is exhibited by 
the Sab-Sc spirals, with $14 < T_{\rm D}^{\rm cold} < 28$\,K. The later 
spirals and
irregulars have a distribution shifted towards colder dust temperatures, with
the BCDs having the coldest dust temperatures. The BCD VCC~655 has the 
coldest dust temperature (12.6\,K) among the galaxies with 3 wavelengths 
detection. Of the BCDs detected only at two wavelengths, 
VCC~848 has the coldest dust temperature, 9.9\,K as an upper limit. The 
median values of the dust
temperatures (Table~4) are in agreement with the early-type spirals having the
warmest median temperature and the BCDs having the coldest median 
temperature.

\begin{table}[htb]
\tablenum{4}
\caption{Median values for the temperatures and mass surface densities within 
each bin of Hubble type}
\begin{tabular}{lll}
\hline
type                   &
$T_{\rm median}$       &
$(M/D^2)_{\rm median}$ \\
                       &
K                      &
M$_{\odot}\times {\rm h}^{-2}\times{\rm arcmin}^{-2}$\\ 
\hline
S0a-Sa & 20.5 & $3.4\times10^4$\\
Sab-Sc & 16.1 & $3.2\times10^5$\\
Sd-Sm  & 18.5 & $2.1\times10^5$\\
Im-BCD & 15.9 & $1.2\times10^6$\\
\hline
\end{tabular}
\end{table}

The distribution of cold dust masses (Fig.~6b) shows a pronounced effect of the
scaling relations, with the Sab-Sc spirals having the largest amount of dust,
as expected for more massive galaxies. However, in the dust mass surface 
density distributions (Fig.~6c) there is a trend for the early spirals to have
small dust masses per unit area and for the BCDs to have the larger 
amounts of dust compared to their optical sizes. The BCDs VCC~655 and VCC~848 
are the most extreme examples. The median of the dust mass surface density 
(Table~4) has the lowest value for the early-type spirals and the largest 
value for the BCDs. 
There is no obvious trend in the distributions of the dust-to-HI mass 
ratio (Fig.~6d) with respect to Hubble type.

The distributions of FIR luminosities (Fig.~7a) are mainly dominated by the
scaling effects, with the more massive galaxies (Sbc-Sc) having the larger FIR
luminosities. Only after normalization to the K$^{\prime}$-band magnitudes 
(Fig.~7d)
does it become obvious that there is a trend for the early-type spirals to have
intrinsically low total FIR luminosities; this is mainly 
attributable to the
``one component'' galaxies.  

Within the available statistics there was no evidence for a strong segregation
between cluster periphery and cluster core galaxies within morphological
classes. The only exception might be the tendency for cluster core early type
spirals (which are known to have the most extreme HI deficiencies) to be
lacking cold dust components.

\section{Correlations with indicators of SFR}

\subsection{The FIR-H${\alpha}$ correlation}

The H$\alpha$ equivalent widths (EW) measure the strength of 
the mass-normalized recent ($< 10^8$\,yr) SFR. To study the correlation
between the H$\alpha$ EW and the FIR emission for our sample galaxies we made
use of the data available in the literature, namely 
$\rm H \alpha +$[NII] EW obtained from long-slit (from 3 to 7 arcmin) 
spectroscopy (Kennicutt \& Kent 1983) or from CCD imaging 
(Gavazzi, private communication). Their typical uncertainty is 2\,${\AA}$. 
Fig.~8a and 8b show the normalized FIR luminosities of the warm and 
cold dust components, respectively, versus the H$\alpha$ (+[NII]) equivalent 
widths (EW). The linear correlation found 
for the warm component is consistent with, and a necessary condition for our 
original premise for the constrained fit, namely the 
identification of the warm component with localized HII regions. 

To better understand the trends in the correlations we consider below the 
bolometric energy budget as a function of the recent SFR, in the case of a 3
component model consisting of: locally heated dust in star-forming
complexes, 
diffuse dust heated by the non-ionizing UV photons, and
diffuse dust heated by the optical photons. Such a model does not take into
account  the existence of cold dust associated with clumps 
(quiescent or associated with molecular clouds), which may need a separate 
treatment. However, if the cold
dust is only in a diffuse component, the relation between SFR and the total 
FIR energy output can be derived from the following equation:
\begin{eqnarray}
L_{\rm FIR}^{\rm tot}=L_{\rm FIR}^{\rm HII}+
L_{\rm FIR}^{\rm UV}+L_{\rm FIR}^{\rm opt}
\end{eqnarray}
where $L_{\rm FIR}^{\rm tot}$ is the total FIR luminosity emitted by the
galaxy, $L_{\rm FIR}^{\rm HII}$ is the FIR luminosity emitted by the 
HII regions and star forming complexes, $L_{\rm FIR}^{\rm UV}$ is the FIR 
luminosity of the diffuse dust component heated by the non-ionizing 
UV photons and $L_{\rm FIR}^{\rm opt}$ is the FIR luminosity of the diffuse 
dust component heated by the optical photons. The equation can be further 
expressed in terms of SFR:
\begin{eqnarray}
L_{\rm FIR}^{\rm tot}= SFR \times (L_0 \times F + L_x \times X) + 
SFR \times L_0 \times (1-F) \times G_{uv}+L_{\rm FIR}^{\rm opt}
\end{eqnarray}
where SFR is the present-day star formation rate in 
${\rm M}_{\sun}/{\rm yr}$, $L_0$ and $L_x$ are the non-ionizing and the ionizing
 UV bolometric luminosities of a young stellar population corresponding to 
$SFR = 1\,{\rm M}_{\sun}$/yr (which can be derived from population synthesis 
models), F and X are the fractions of non-ionizing and ionizing UV emission
that is absorbed by dust locally within star forming complexes,  and the factor
$G_{uv}$  is the probability that a non-ionizing UV photon escaping from the
star formation complexes will be absorbed by dust in the diffuse interstellar 
medium. The ionizing UV is thought by most authors to be mainly locally
absorbed by gas in the HII regions. Its contribution to
$L_{\rm FIR}^{\rm HII}$ is in any case small, and its effect on the
heating of dust in the diffuse ISM can be totally neglected.

The warm dust component from our fitting procedure was identified with the dust
locally heated within the HII regions, such that
$L_{\rm FIR}^{\rm warm}\simeq L_{\rm FIR}^{\rm HII}$. 
In this case we indeed expect a linear correlation with the SFR and thus 
with the EW, 
\begin{eqnarray}
L_{\rm FIR}^{\rm warm}\simeq SFR\times (L_0\times F + L_x\times X)
\end{eqnarray}
The scatter in the correlation of Fig.~8a is only to be expected due to the 
possible contribution of the diffuse component to the 60\,${\mu}$m 
band via stochastic emission, heating by the old stellar population
(see Popescu et al. 2000), and the likely variation in HII region dust 
temperatures within and between 
galaxies (e.g. due to the effect of metallicity on the hardness of the stellar 
photons and/or  due to effects of star-burst age). To this we should add the
uncertainty of $\sim0.3$ dex in the value of the 
H${\alpha}$ EW (due to measurement errors in determining the continuum emission
as well as to the [NII] contamination), which can also contribute to the 
scatter.  

If the cold dust component is identified with diffuse dust heated by the
interstellar radiation field, then:
\begin{eqnarray}
L_{\rm FIR}^{\rm cold}\simeq
SFR \times L_0 \times (1-F) \times G_{uv}+L_{\rm FIR}^{\rm opt}
\end{eqnarray}

Fig.~8b suggests that there is also a correlation, though non-linear, and 
with a larger scatter, for the cold component. The correlation coefficient 
is 0.75 if we consider all the data points or 0.83 if we exclude 
the extreme point above the correlation (at log EW(H${\alpha}$)=0.7). The
excluded galaxy is a BCD, and we have 
already mentioned the unusual properties of these galaxies.
The correlation
itself suggests that the cold dust component is predominantly heated by
the non-ionizing UV radiation produced by the young stellar population,
consistent with the model predictions of Popescu et al. (2000) and Misiriotis
et al. (2001). To explain the non-linearity in the correlation based on the
model described above we can invoke either a variation of the $G_{uv}$ factor
with H${\alpha}$ EW or a strong influence of  
$L_{\rm FIR}^{\rm opt}$ for small H${\alpha}$ EW. The F factor is taken to be
invariant within our sample, an assumption 
validated by the linear correlation obtained in Fig.~8a. The first 
possibility would imply that more active galaxies have lower $G_{uv}$ 
factors (less optically thick 
disks). This seems unlikely, as starburst galaxies are known to be
more optically thick systems compared to normal galaxies. The second
possibility simply states that more quiescent galaxies should have a 
higher contribution from the optical photons in heating the dust. Scatter 
in the correlation is again to be expected due to the varying contribution of 
the old 
stellar population in heating the dust and to some varying degree of 
clumpiness of the interstellar medium within galaxies.

\subsection{The FIR-radio correlation}

Another important correlation with indicators of SFR is the well known
FIR-radio correlation. Discovered during the IRAS mission, the FIR-radio
correlation was probed not only in terms of absolute fluxes 
(e.g. Helou, Soifer \& Rowan-Robinson 1985, de Jong et al. 1985, 
Wunderlich, Wielebinski \& Klein 1987) but also in flux per unit
galactic mass (Xu et al. 1994). Basically the correlation holds for the 
integrated emission. The link
is qualitatively given by the grain heating associated with
the appearance of massive stars and the acceleration of relativistic
particles in their eventual SN explosions. However, the tightness of the
correlation appears to constrain the characteristics of late-type galaxies
to rather more specific properties: an optical thickness of the disk that
is of order unity for non-ionizing stellar UV radiation (Xu 1990), a
magnetic field energy density that is about equal to that of the radiation
field over more than an order of magnitude in dynamic range, and radiative
energy losses of the synchrotron electrons that exceed escape losses, so that
the galaxies act for both energy inputs as calorimeters (V\"olk 1989, 
Lisenfeld, V\"olk \& Xu 1996). 

Previous studies of the FIR-radio correlation were confined to the IRAS fluxes,
and to FIR luminosites obtained by extrapolating the IRAS fluxes. Our new
ISOPHOT data, and in particular our finding of a cold dust component present
within all morphological classes naturally raises the question of whether the 
FIR luminosity associated with the cold component is also correlated with the
radio emission. Popescu et al. (2000) have shown that in NGC~891 there is a
relative increase of the contribution of the UV photons to the heating of the
diffuse interstellar dust with increasing FIR
wavelengths, such that in the sub-mm regime the dust emission is mainly powered
by the UV photons ($\sim 60\%$). If this is the case one would also predict a 
tight correlation for the FIR emission of the cold dust component with the 
radio emission, since the stars mainly responsible for the heating of the cold
dust also give rise to most of the radio emission. 
Fig.~9a, b show a good correlation between the FIR luminosities of both the 
warm and cold dust components, and  the NVSS (NRAO VLA Sky 
Survey) radio luminosities at 1.4\,GHz taken from Gavazzi \& Boselli (1999). 
The interpretation of the correlations is limited by the
poor statistics, since only 14 galaxies from our sample have NVSS
detections. Nevertheless, some interesting features are apparent. 
First of all, the prediction regarding the validity of 
the FIR-radio correlation for the cold dust component is confirmed. 
Secondly, there is a clear non-linearity in both correlations. 

The non-linear warm FIR-radio correlation plotted in Fig.~9a  
($\rm FIR/radio<1$ and with a correlation coefficient 0.85), is similar
(within the poor statistics) to that obtained by 
Xu, Lisenfeld \& V\"olk (1994). However, the ``warm'' and  ``cold'' FIR 
components from Xu et al., as  derived from the IRAS 
observations, are not to be identified with our 
``warm'' and ``cold'' dust components, despite a common physical
justification. The non-linearity of the correlation is an interesting effect,
which at present is difficult to explain. Because of the weak dependence of 
the radio synchrotron intensity on the B field predicted by the original 
calorimeter theory, consistent with observed radio spectral indices
(Lisenfeld \& V\"olk 2000), the non-linearity in the warm FIR-radio 
correlation should be small. The 
scatter in the warm FIR-radio correlation has probably several components: 
the intrinsic scatter of the FIR-radio 
correlation (see Xu, Lisenfeld \& V\"olk 1994), and, as for 
the case of the warm-FIR H$\alpha$ correlation, components due to 
effects of stochastic heating of dust in low density radiation fields and to 
optical heating of dust in high density radiation fields, in optically thick 
regions in the very central part of galaxies.

The non-linear cold FIR-radio correlation plotted in Fig.~9b
($\rm FIR/radio<1$ but closer to slope unity and with a correlation
coefficient 0.85) is again similar to that obtained by 
Xu, Lisenfeld \& V\"olk (1994).  Unlike the case of the warm FIR-radio
correlation, the non-linearity of the cold FIR-radio correlation can be 
explained by invoking a higher relative
contribution to the heating of diffuse dust by the old stellar population in
galaxies with lower present day SFR (Xu et al. 1994). The scatter of the 
correlation has again to be viewed in terms of several components, where one 
component is attributable to the contribution of varying proportions of 
optical photons in heating the dust.

\section{Discussion}

The statistical analysis of the integrated FIR properties of our Virgo sample
unambiguously shows the existence of a cold dust component present in all
morphological classes of late-type galaxies, from early spirals to irregulars
and BCDs.
Here we consider the implications of this analysis for the nature of the 
cold dust component.

A basic question arising from our analysis concerns
the location and morphology of the cold dust component. One argument in favor 
of a diffuse origin for this component is the lack of a cold 
component in some Virgo cluster core early type galaxies with HI 
deficiencies. Though statistics are poor, this may point towards a diffuse 
dust component 
which is absent because of stripping. Another hint for a diffuse origin of the
cold component is the non-linear correlation between the normalized luminosity
of the cold dust component and the H${\alpha}$ EW seen in Fig.~8b, which was
interpreted in terms of a diffuse dust component heated by both UV and
optical photons, with the main contribution coming from the UV
radiation (as predicted by Popescu et al. 2000, Misiriotis et al. 2001). 
Finally, the cold FIR-radio correlation is again consistent with the 
predictions of the Popescu et al. (2000) model, where the bulk of the cold FIR
emission arises from diffuse dust heated by UV photons.

A fundamental issue is the amount of dust in galaxies. We find large amounts
of dust in galaxies, significantly higher than derived from IRAS data
alone. The missing dust, not seen by IRAS, is present within all 
morphological types of galaxies, from early spirals to irregulars and BCDs. 
However, what we can really measure is dust opacity to starlight, and what we 
derive as dust mass depends on several assumptions. It is not only 
the fact that SED fits are ambiguous with respect to a zero
point (see Sect.~4), but also the possibility that some dust might be 
located in compact optically thick sources.  In
such cases dust may have different optical properties, and in particular 
ices may be present. These have stronger mass-normalized absorption 
coefficients in the FIR-submm regime (e.g. Preibisch et. al. 1993), which 
would reduce the implied dust-to-gas ratios for these regions. 

Perhaps the most intriguing result of this investigation 
are the masses and temperatures of the cold dust derived for
the BCD galaxies in our sample. 
These systems are clearly differentiated from the spirals, having the 
highest dust mass surface densities (normalized to optical size), 
and the lowest dust temperatures (ranging down to less than 10\,K ). 
This is not the expected behavior for
galaxies whose FIR emission is dominated by dust heated locally
in HII regions, where one would anticipate temperatures of 30\,K 
or more. This was the a priori expectation in particular 
for the BCDs, and was the standard interpretation for the IRAS results 
obtained for these systems (see for example Hoffman et al. 1989,
Helou et al. 1988, Melisse \& Israel 1994, who found that the 
60/100\,${\mu}$m colors of BCDs - including examples from the Virgo cluster - 
were clearly warmer than for spirals). Our findings indicate a cold dust
component supplementary to the warm dust component detected by IRAS.
This in itself is not surprising, as it is to be expected that
cold dust associated with the star formation regions (SFRs) 
should be present at some level in association with the 
molecular component detected in CO.
The particularly unexpected aspect of this is rather that the luminosity
of the cold dust component should dominate over that of the warm dust, as 
shown by the failure to detect systems (which are clearly seen at 170 
micron), in one or both of the ISOPHOT 60 and 100\,${\mu}$m bands.\footnote
{The integrated Blue magnitudes of the BCDs in our sample 
are fainter than the BCDs detected by IRAS by typically two magnitudes.}
An equally strong observational constraint is given by the ratio
of observed luminosity of the cold dust component compared with the 
observed (i.e. not corrected for intrinsic absorption) luminosity in 
B-band.\footnote{This was estimated by taking the $B_T$ magnitudes from 
Binggeli, Sandage and Tammann (1985) and converting to fluxes using 
conversion factors tabulated in Matthews \& Sandage (1963).} This ranges from 
6 to 60 percent over the sample of BCDs.

Here we qualitatively discuss the nature of the cold 
dust emission detected by ISO in Virgo Im and BCD galaxies, 
considering in general terms the origin and location of the grains, and 
their heating. 
For photon-heated dust, low grain temperatures might be reached in dense 
clouds opaque to the ambient radiation fields in the galaxies, 
provided the clouds were quiescent ( i.e. not harboring
star forming regions which would give rise to a warm FIR emission component
dominating the bolometric output of the clouds.) 
However, clouds with filling factors of perhaps a few percent would only be 
expected to intercept and reradiate
a corresponding few percent of the UV-optical output of the
galaxy. By contrast, perhaps a few tens of percent
of the UV-optical output of these systems appears in the cold
dust component.

Alternatively, the grains could
be distributed in a diffuse component sufficiently extended that
the interstellar radiation field (ISRF) would be weak enough to lead to grain 
temperatures of order 10\,K or less. Such grains would have to be
distributed over dimensions of order 10\,kpc around the central
star-forming area in the BCDs (of extent typically of order a kpc). The
grains would be embedded in the surrounding intergalactic medium (IGM), most
probably in the putative protogalactic cloud from which the galaxy formed.
Such an explanation might explain the extreme cold dust surface densities 
for Im and BCD galaxies, which would be reduced to values more compatible 
with spirals if the surface area normalization were to be with respect to a 
radius compatible with the cold dust temperature rather than with respect
to the B-band extent of the galaxies. Furthermore, there is
evidence that the 170${\mu}$m emission is resolved in two BCDs, VCC~1 and 
VCC~848 (see Fig.~5 of Tuffs et al. 2002).
However, if such grains are photon-heated, significant optical
depths in the diffuse component would be required to convert
substantial fractions of the UV-optical output into FIR photons,
as observed in some objects. In their sample of 13 field BCDs, Hunter \&
Hoffman (1999) found E(B-V) to be distributed in the range $0.00 - 0.53$,
with evidence for optical colors being affected by a diffuse dust
component.

Another possibility to account for FIR emission of the observed
luminosity, color and extent would be to invoke collisional, rather
than radiative heating for the extended cold dust emission component.
In particular, BCD galaxies are thought to undergo
sporadic episodes of star formation activity (Mas-Hesse \& Kunth 1999), 
giving rise to a galactic wind which interacts with the protogalactic cloud. 
This creates a wind bubble containing shocked coronal swept-up gas, in
which any embedded grains will be collisionally heated. If collisional
grain heating were to constitute the dominant cooling mechanism for the
shocked swept up IGM, the observed FIR luminosity and 
color of the cold dust emission from the Virgo BCD galaxies could be
explained. 

A detailed description of the circumstances under which such an
efficient conversion of mechanical wind luminosity into 
FIR radiation occurs, is beyond the scope of this paper.  The general 
analytical 
solutions of Weaver et al. (1977) can be applied to the gas density and 
temperature profiles of a spherical wind bubble, to explore
the conditions under which we may expect the internal energy of the
coronal gas in the shocked swept up IGM region to be converted 
predominantly into FIR radiation, as required by the observations.
In essence, this can only occur if sufficient dust is present in the
shocked IGM that the timescale for gas cooling through inelastic
collisions with grains is both shorter than the timescale for
cooling through line emission, as well as being less than or
comparable to the dynamical timescale.
This scenario would have interesting implications for the
interpretation of the integrated FIR emission of dwarf star-forming galaxies
in the distant universe, implying that some fraction
of the FIR luminosity had a mechanical, rather than photon-powered
origin. One immediate observational consequence would be a larger 
color ratio $L_{\rm FIR}/L_{\rm UV-opt}$ for the ensemble of
cosmologically distant dwarf star-forming systems than would be expected
on the basis on photon heating alone, as we have directly
observed for our sample of Virgo BCD systems.

A rather different scenario explaining the high dust mass and dust surface
densities would come back to a spatially extended grain population that is
asymmetrically distributed around the optical-UV emitting central star-forming
region. This could be a (flaring) disk of accreting material that feeds the
star forming center. BCDs would then be those members of such asymmetric
systems seen under favorable aspect angle for observation of the central
optical-UV emission. Systems seen edge-on would be FIR-dominated by photon
heating and possibly not yet detected as a group in the visual range. They
would only show up in the FIR and sub-mm range, where the only major blind
search up to now is the ISOPHOT 170\,${\mu}$m serendipity survey. It will be
interesting to search for such objects in the serendipity sample. In the
extreme case BCDs - or a significant sub-class of them - would be dwarf
galaxies that undergo a massive gas/dust accreting phase that makes them (at
least sporadically) bright optical-UV sources. Collisional dust heating might
occur in addition as a consequence of dynamical effects like wind interactions
in the broader environment, as described above.
 
\section{Summary}

Based on observations taken in the P32 mode of the ISOPHOT instrument on
board ISO we have statistically analyzed the integrated FIR properties
of a complete volume- and luminosity-limited sample of late-type Virgo
Cluster galaxies. For the first time, data taken in this observing mode
could be corrected for the complex non-linear response of the detectors,
allowing robust integrated photometry to be extracted, and over the
entire optical extent of the galaxies, including the outer regions of
the galactic disks. We demonstrate the existence of a cold dust component 
present within all the morphological classes observed, which range from S0a 
to Im and BCD. This cold component, which was not previously seen by IRAS,
was analyzed by fitting the data with a superposition of two
modified black-body functions of form ${\nu}^2\,{\rm B}_{\nu}$,
physically identified with a localized warm dust emission component
associated with HII regions (whose temperature was constrained
to be 47\,K), and a diffuse emission component of cold dust.
The fits imply a revision of the masses and temperatures of
dust in galaxies.

The main results are summarized as follows:

\begin{itemize}
\item The dust masses should be raised by 
factors of $6-13$ from the previous IRAS determinations, with even larger 
factors for certain individual galaxies. The temperature of the cold dust is 
found to be generally $8-10$\,K lower than the IRAS temperatures, again with 
individual galaxies having even lower temperatures. 
\item The temperatures of the cold dust component have a tendency to become 
colder for the later types. The early-type spirals have a  
distribution in cold dust temperatures shifted towards higher values, 
with the coldest and warmest temperatures of 17.7\,K  and 33.4\,K. 
The later spirals and irregulars have a distribution 
shifted towards lower dust temperatures, with the BCDs having the coldest 
dust temperatures (ranging down to less 
than 10\,K).
\item There is a trend for the early spirals to have small dust 
masses per unit area and for the BCDs to have the largest amounts of dust
normalized to their optical sizes.
\item The BCD galaxies were found to have the
highest dust mass surface densities (normalized to
optical area) and the coldest dust
temperatures of the galaxies in the sample.
This is a particularly unexpected result,
since the IRAS observations of BCDs could be accounted for
in terms of dust heated locally in HII regions,
with temperatures of 30\,K or more. Two scenarios invoking
collisionally or photon-heated emission from grains originating
in the surrounding intergalactic medium are proposed to
qualitatively account
for the FIR and optical extinction characteristics of BCDs.
In the one scenario, grains are swept up from a surrounding protogalactic
cloud and heated collisionally in an optically thin wind bubble blown
from the BCD. In the other, the grains are taken to be photon-heated
in an optically thick disk surrounding the optical galaxy. The disk is
indicative of a  massive gas/dust accreting phase which makes dwarf
galaxies sporadically bright optical-UV sources when viewed out of
the equatorial plane of the disk. Elements of both scenarios
may apply to real-life BCDs.
\item A good linear correlation is found between the ``warm FIR'' luminosities
and the H${\alpha}$ EW. This is in agreement with the assumptions of our
constraint SED fit, that the ``warm'' component is mainly associated with 
dust locally heated within star forming complexes. We also found a good
but non-linear correlation between the ``cold FIR'' luminosities and the 
H${\alpha}$ EW. The correlation itself confirms the predictions of the Popescu
et al. (2000) model, where the emission at FIR-sub-mm wavelengths is mainly due
to the diffuse UV photons.
\item A non-linear correlation is found between the ``warm'' FIR
luminosities and the NVSS radio luminosities at 1.4\,GHz.  A good 
FIR-radio correlation was also found for the cold dust component, 
suggesting that the stars mainly responsible for
heating the cold dust are the massive progenitors of supernovae
whose remnants may be the dominant sources of cosmic ray electrons.
Our findings are the first ones to
test the FIR-radio correlation using the FIR luminosities associated with the
cold dust component. 
\end{itemize}

\vspace{1cm}

This research has made use of the NASA/IPAC Extragalactic
Database (NED) which is operated by the Jet Propulsion Laboratory,
California Institute of Technology, under contract
with the National Aeronautics and Space Administration.

\begin{figure}[htp]
\subfigure[]{
\includegraphics[scale=0.6]{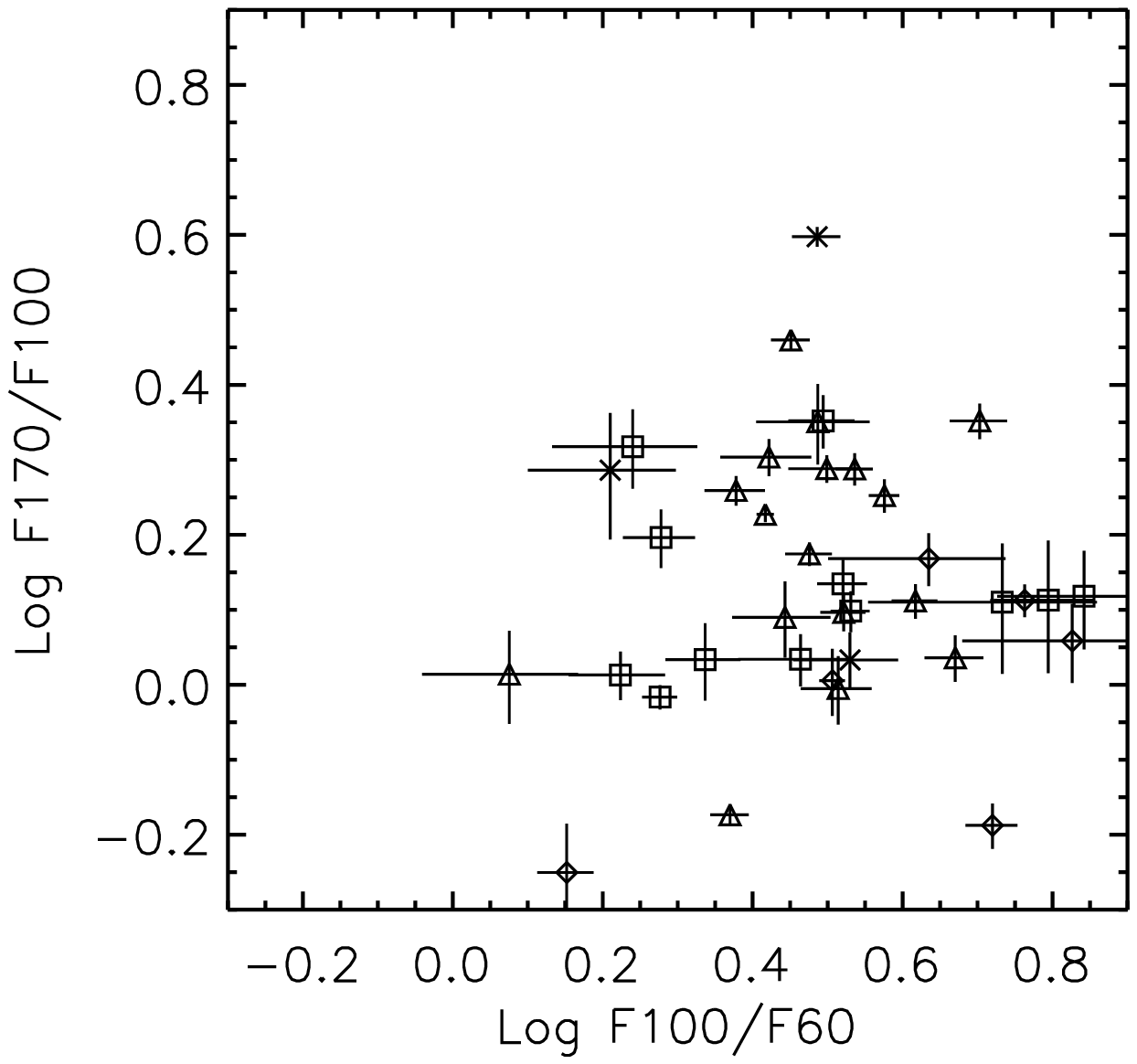}}
\subfigure[]{
\includegraphics[scale=0.6]{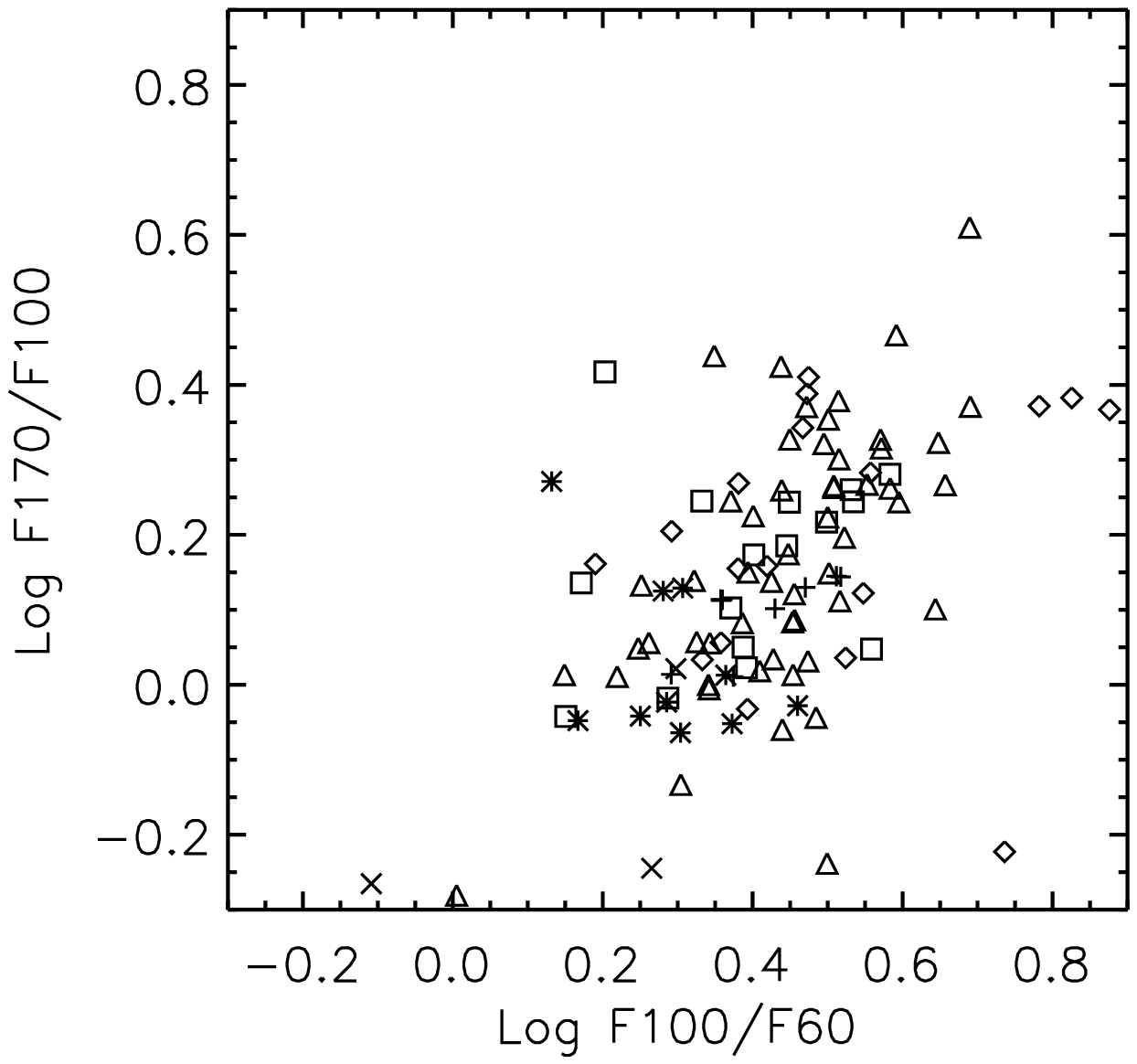}}
\caption{a) The color-color plot of our sample galaxies with detections at all
three wavelengths. The 60 and 100\,${\mu}$m flux densities have been 
converted to the IRAS flux scale. Random uncertainties in the color ratios 
are plotted as bars (half length $1\,\sigma$). b) The color-color plot from 
the ``serendipity sample'' (Stickel et al. 2000). Different Hubble types are 
plotted as follows: diamonds: S0-Sa, triangles: Sb-Sc, squares: Sd-Sm, 
crosses: Im-BCD, plus-signs: S galaxies, stars: unclassified. The last two 
symbols refer to the ``serendipity sample''.}
\end{figure}
 
\begin{figure}
\plotfiddle{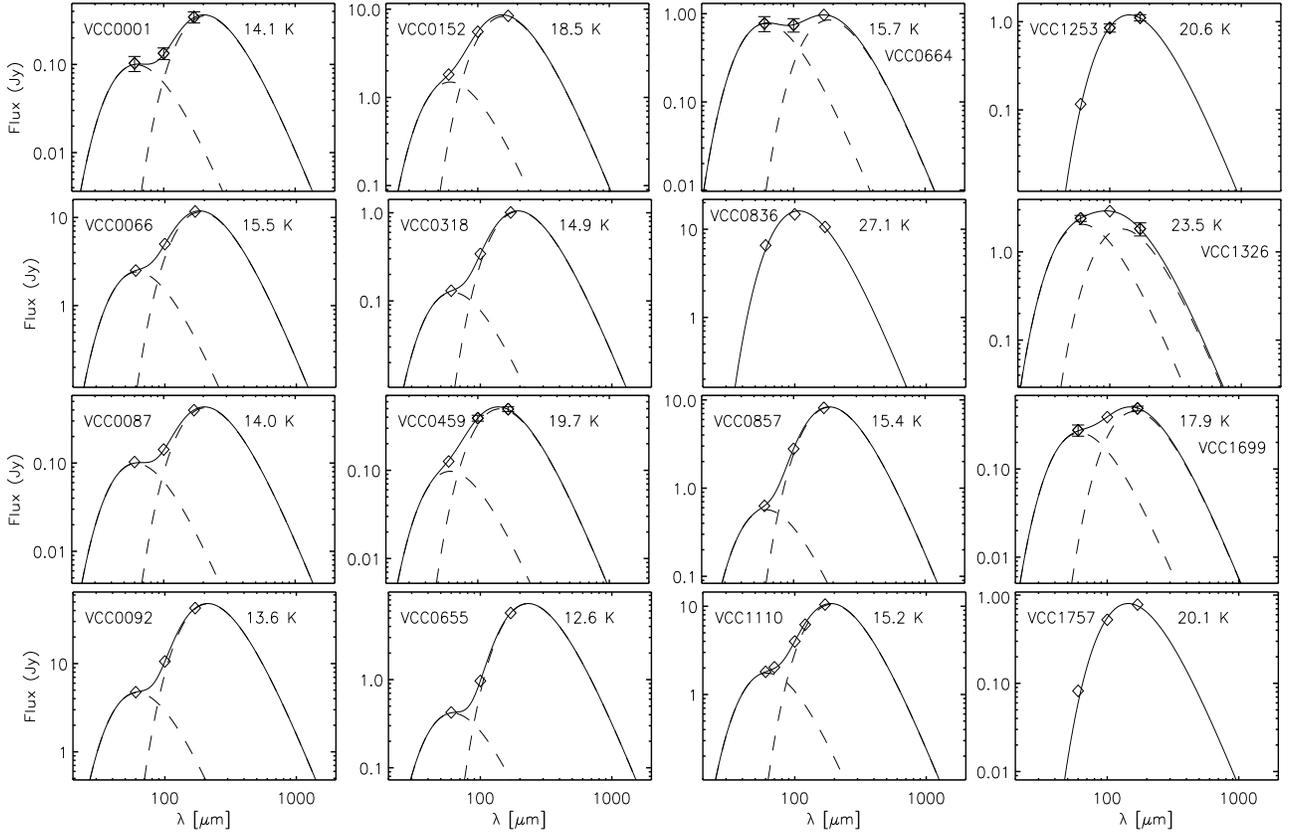}{3.in}{0.}{92.}{100.}{-265.}{-400.}
\caption{Examples of FIR SEDs from our sample galaxies. The color corrected 
flux densities at 60, 100 and 170\,${\mu}$m are plotted together with their 
associated error bars. One galaxy, VCC~1110 has additional measurements at 
70 and 120\,${\mu}$m.
The two modified black-body functions which best fitted the data points are 
plotted with dashed-lines. The temperature of the warm component 
is constrained to be 47\,K. The fitted temperature of the cold component is 
marked near each fit. The sum of the two fitting functions is plotted as the
solid line.  Some galaxies (see text) don't show evidence for
two dust components and their SEDs are fitted with single component 
modified black-body functions, plotted as solid lines.}
\end{figure}

\begin{figure}[htp]
\includegraphics[scale=0.6]{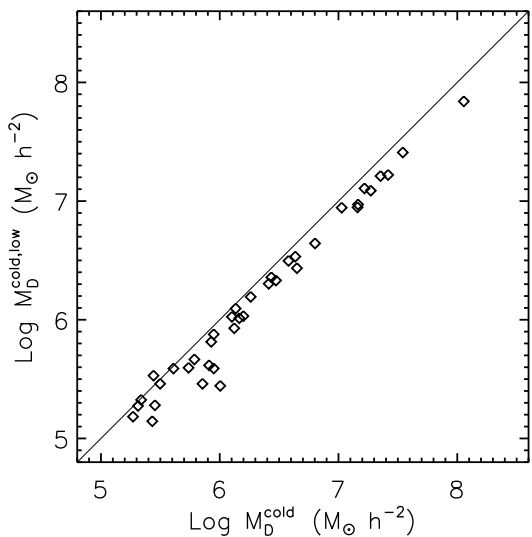}
\caption{The correlation between the mass of the ``cold dust'' component
obtained from fitting two modified black-body functions to the 60, 100 and
170\,${\mu}$m flux densities versus the mass of the ``cold dust'' component in
the lower limit case,  obtained from fitting a single modified black-body
function to the 100 and 170\,${\mu}$m flux densities. The solid line represents
the unit slope zero offset line.}
\end{figure}


\begin{figure}[htp]
\subfigure[]{
\includegraphics[scale=0.6]{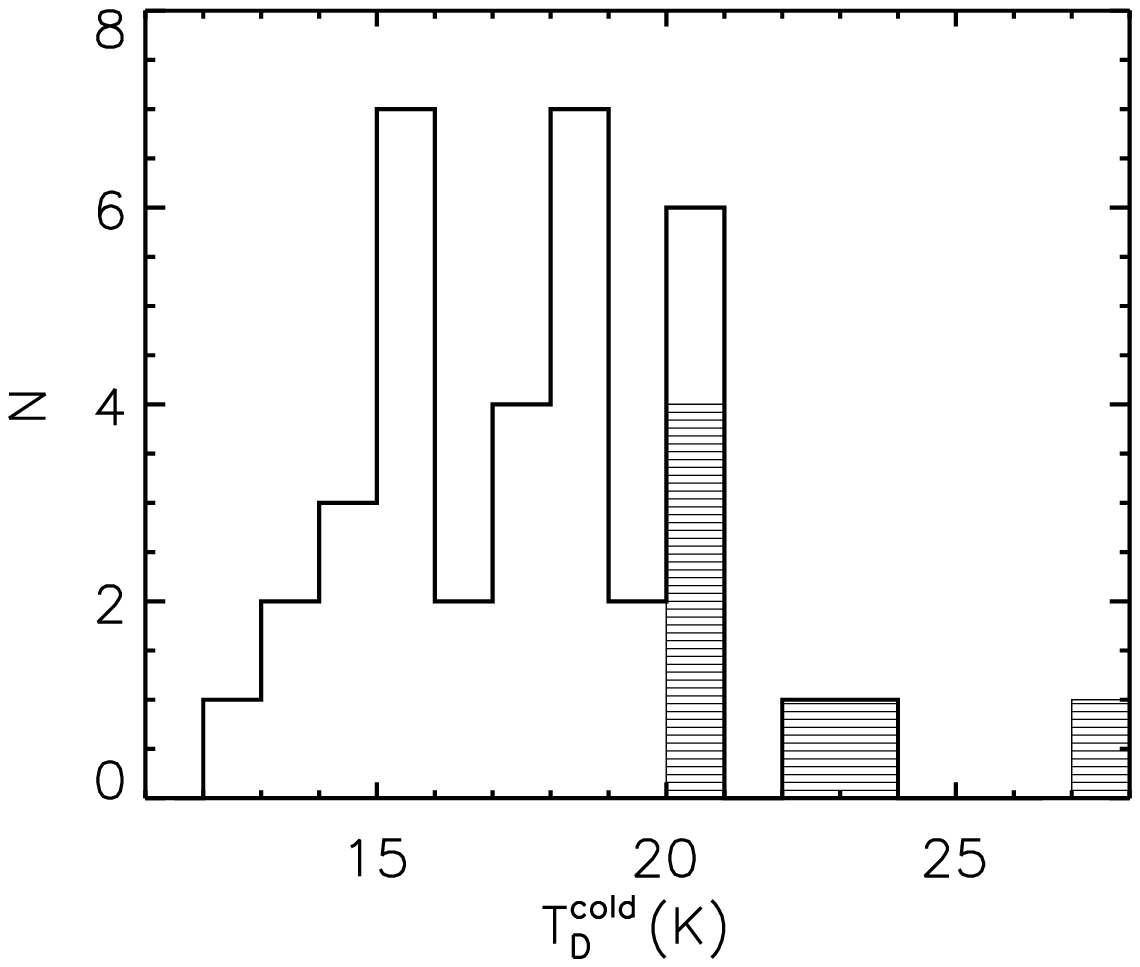}}
\subfigure[]{
\includegraphics[scale=0.6]{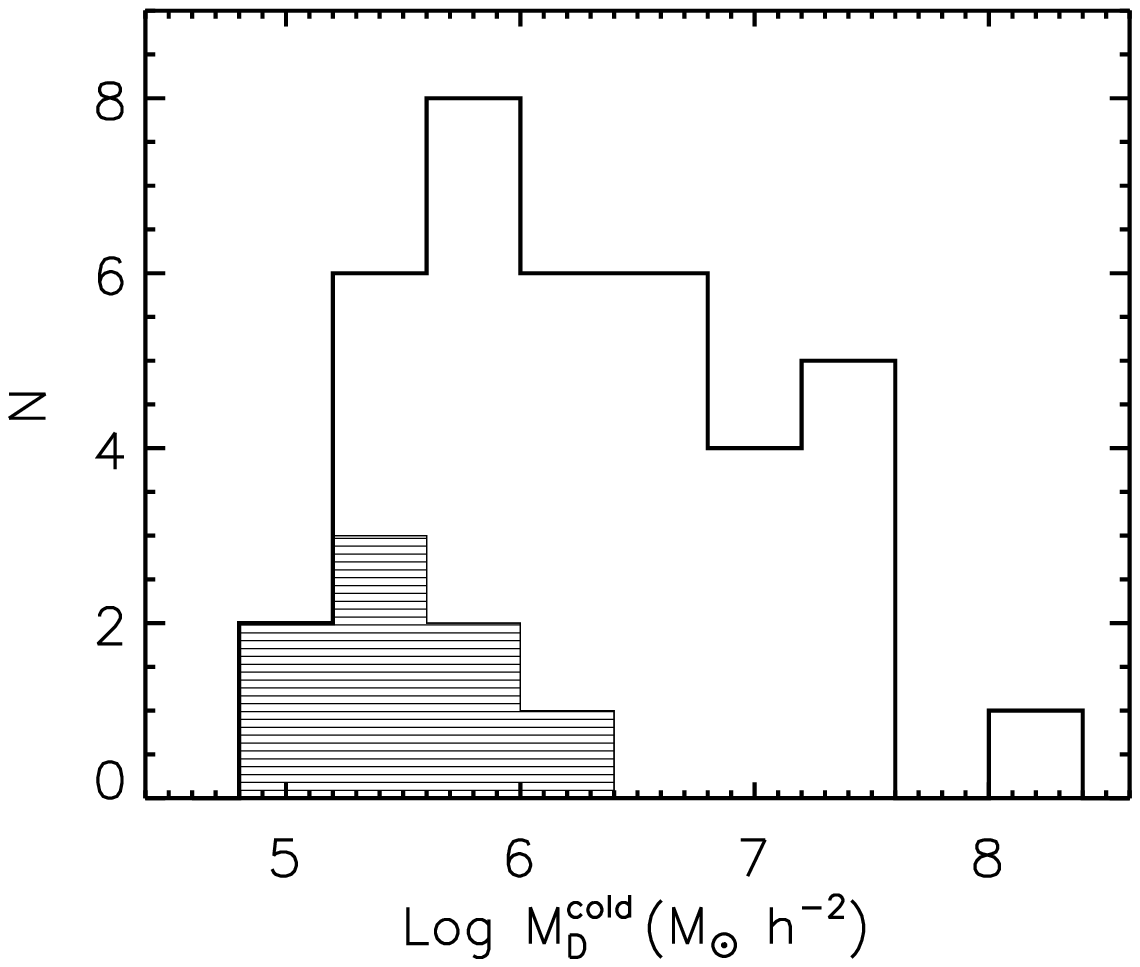}}
\subfigure[]{
\includegraphics[scale=0.6]{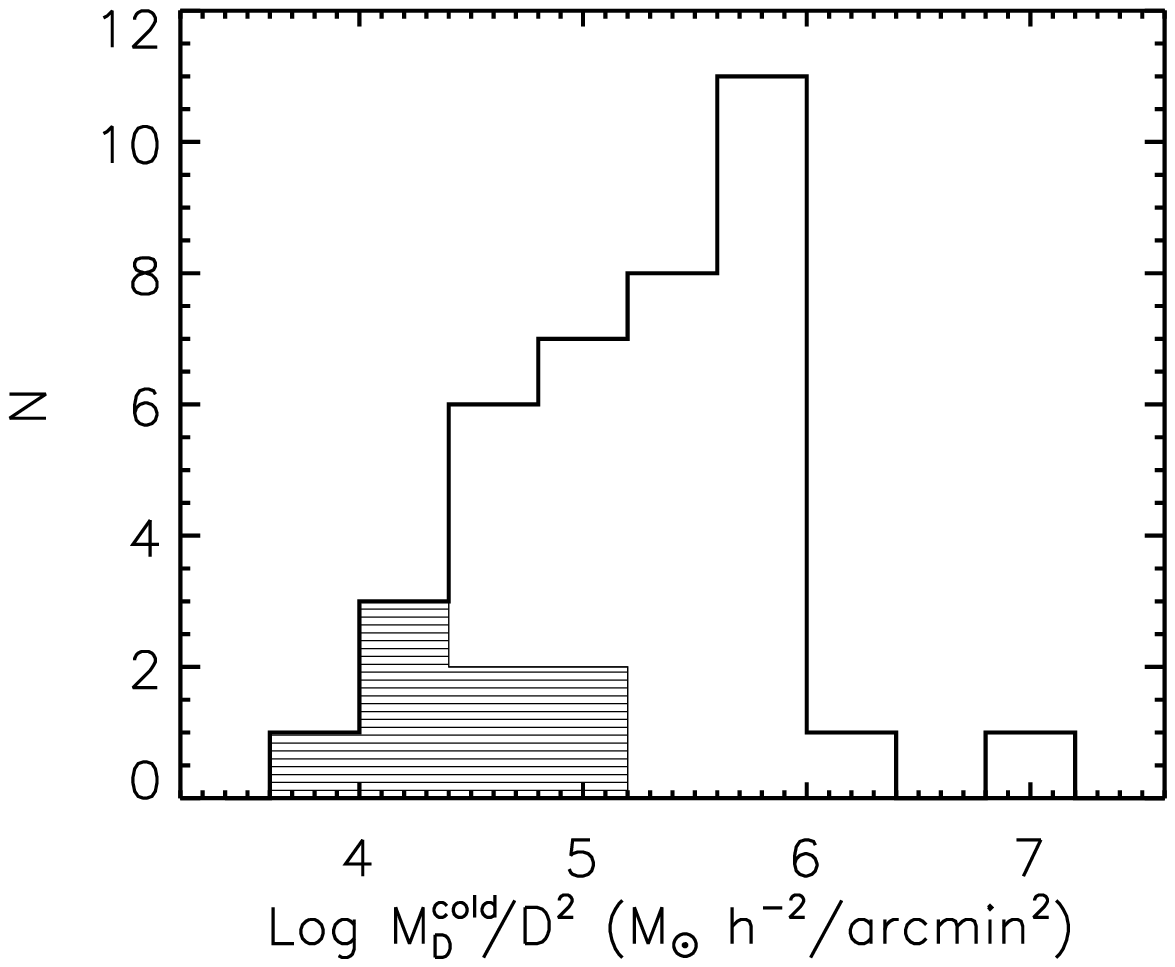}}
\hspace{1.7cm}
\subfigure[]{
\includegraphics[scale=0.6]{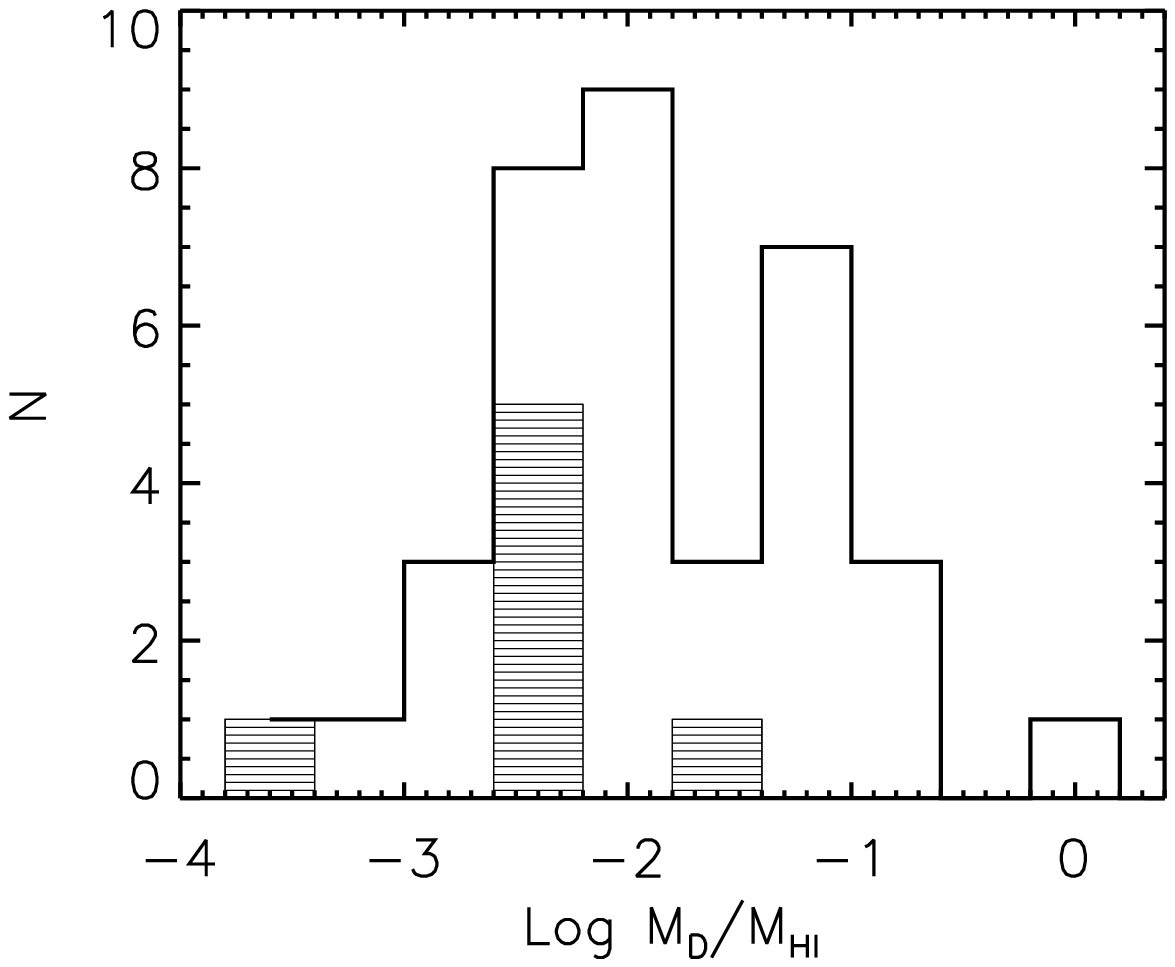}}
\caption{The distribution of a) cold dust temperatures; b) cold dust masses; 
c) cold dust mass surface densities; d) dust-to-HI mass ratio. The hatched 
histograms represent the distributions for the ``one component'' galaxies. The
galaxy with the warmest dust temperature, $T_{\rm D}=34.7$\,K,  was excluded 
from the histogram for display reasons only.}
\end{figure}



\begin{figure}[htp]
\plotfiddle{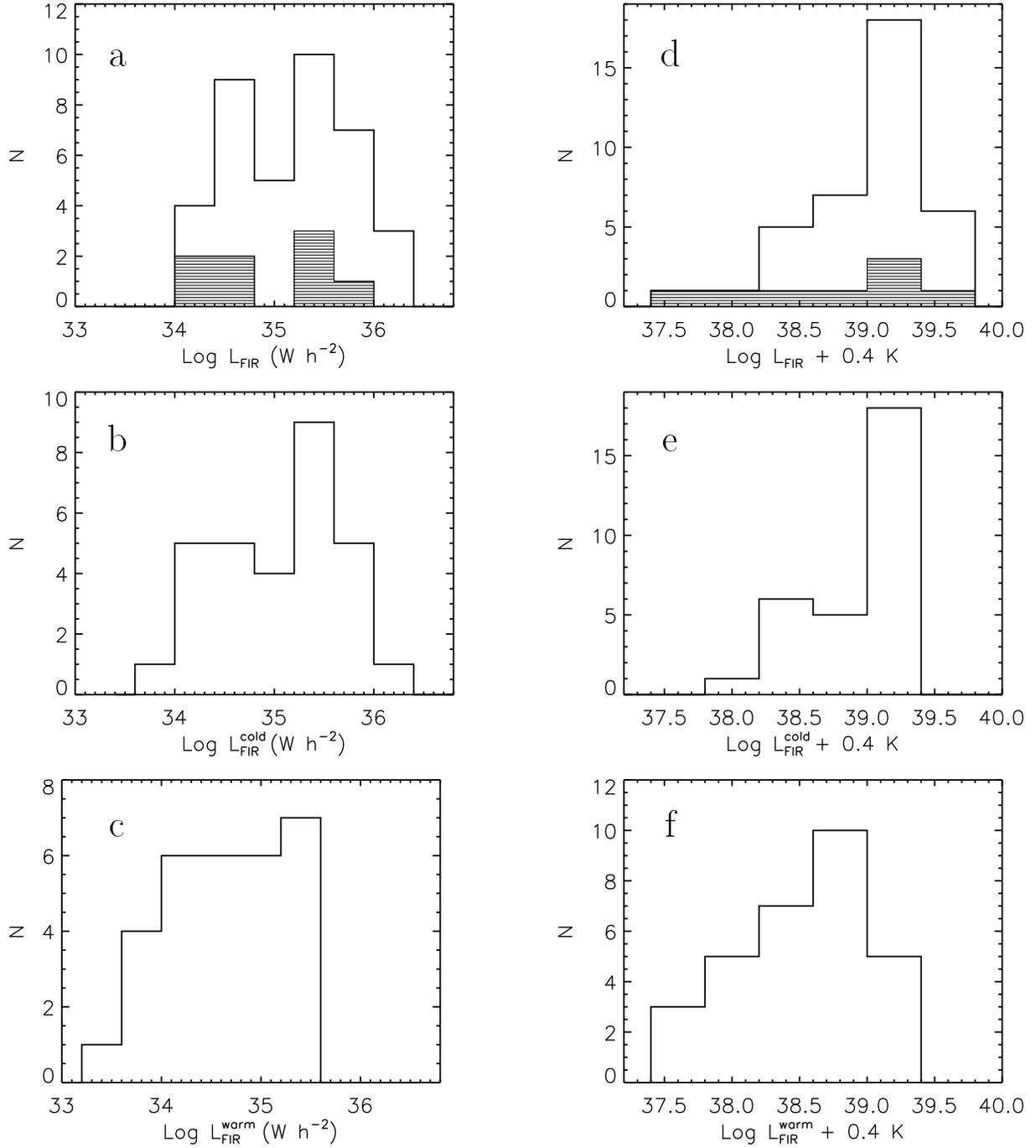}{7.0in}{0.}{100.}{100.}{-310.}{-160.}
\caption{The distribution of a) FIR luminosities; b) FIR luminosities of the 
cold dust component; c) FIR luminosities of the warm  dust component; 
d) normalized FIR luminosities (to the K$^{\prime}$ band magnitude) ; 
e) normalized FIR luminosities of the cold dust component; 
f) normalized FIR luminosities of the warm dust component. The 
hatched histograms represent the distributions for the ``one component'' 
galaxies.}
\end{figure}

\clearpage

\begin{figure}
\plotfiddle{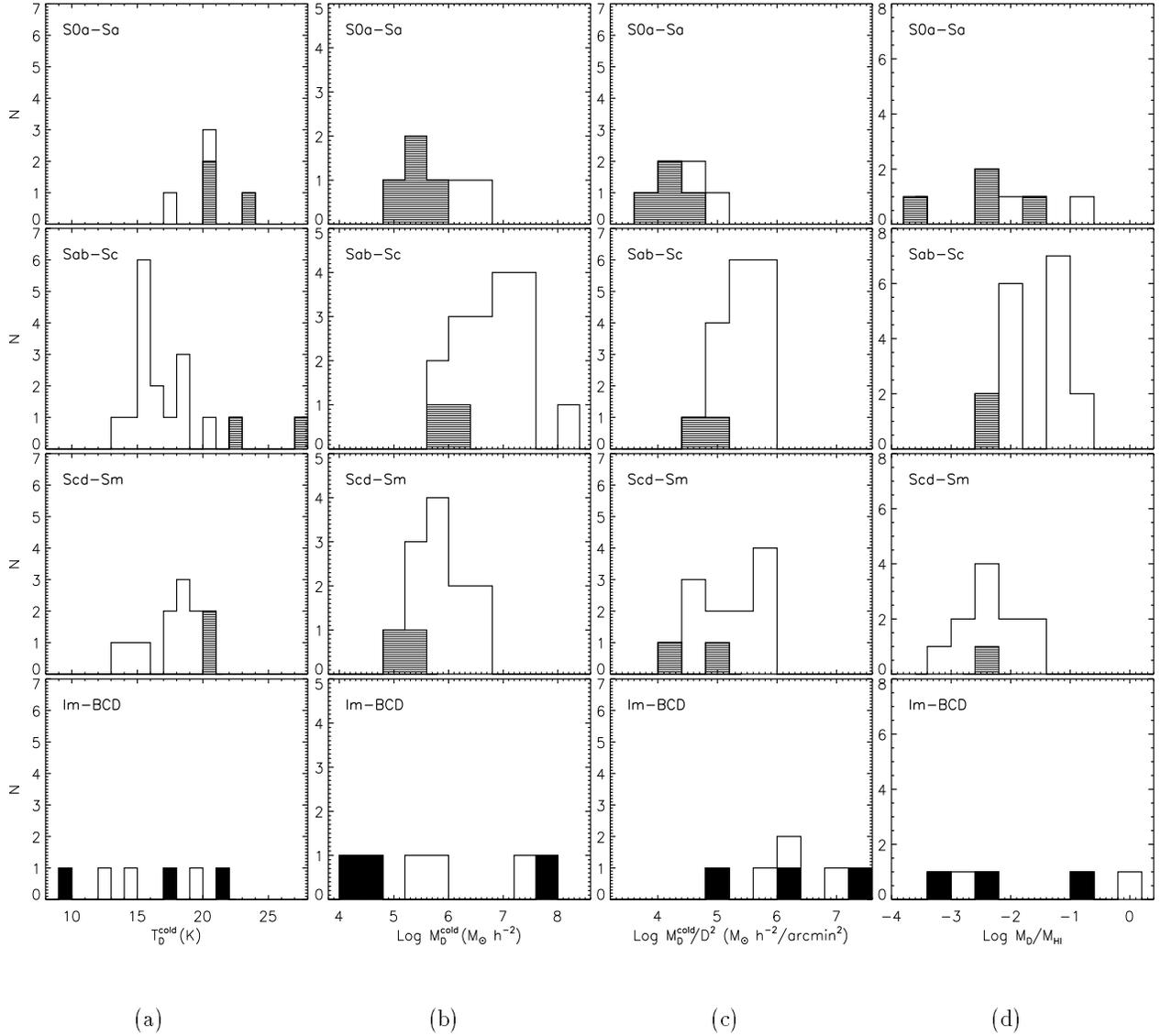}{5.in}{0.}{95.}{100.}{-240.}{-310.}
\caption{The distribution of a) cold dust temperatures 
$T_{\rm D}^{\rm cold}$; b) cold dust masses $M_{\rm D}^{\rm cold}$; 
c) cold dust mass surface densities $M_{\rm D}^{\rm cold}/D^2$; 
d) dust-to-HI mass ratio for different Hubble types. The hatched histograms 
represent the distributions for the ``one component'' galaxies. The filled 
histograms represent the distributions for the galaxies with detections only 
at two wavelengths (100 and 170\,${\mu}$m). For the latter cases the dust 
temperatures are only upper limits and the dust masses are only lower
limits. The galaxy with the warmest dust temperature, 33.4\,K, is not plotted 
in the histogram (panel a, for S0a-Sa) for display reasons only.}
\end{figure}

\clearpage

\begin{figure}
\plotfiddle{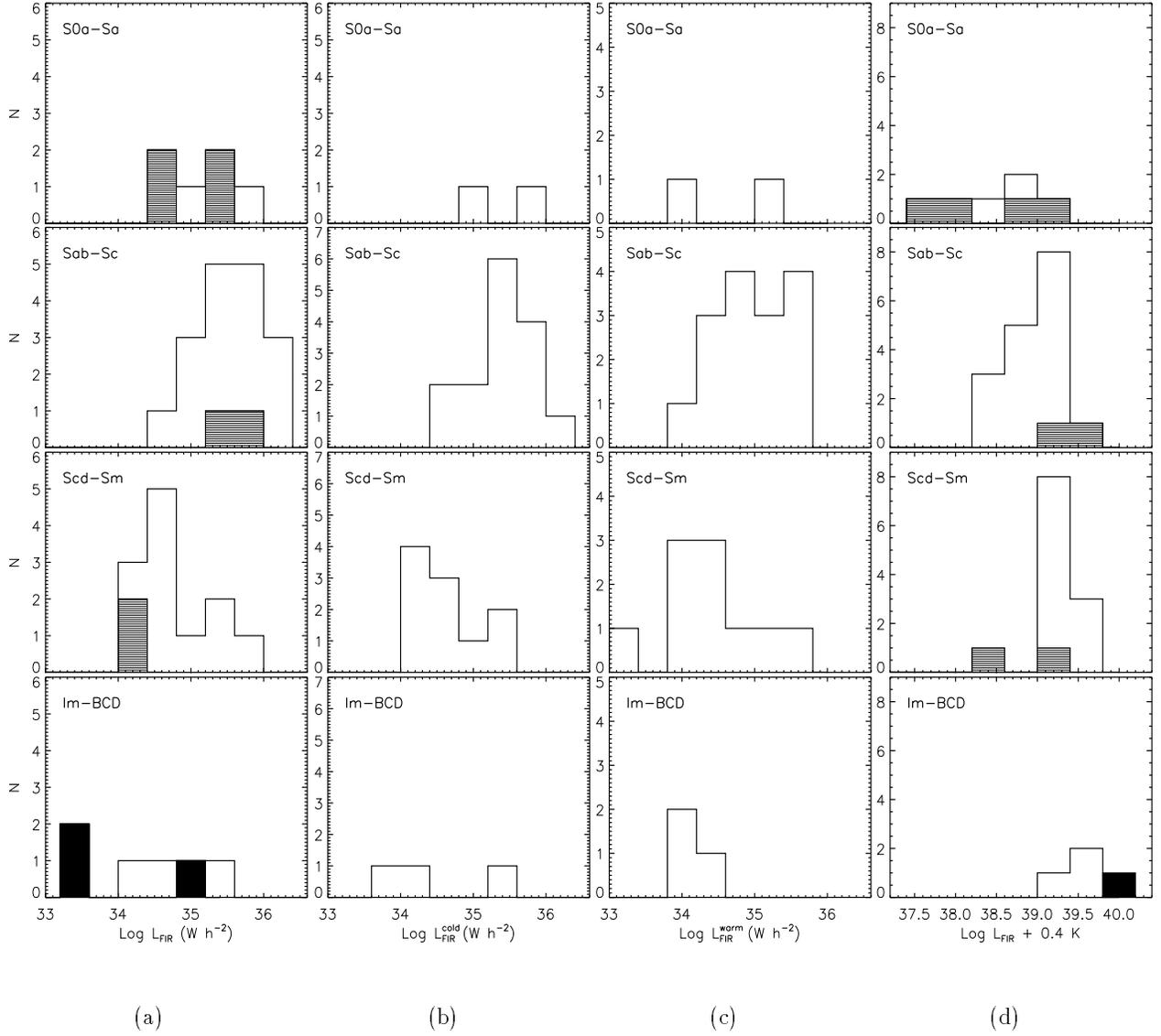}{5.in}{0.}{95.}{100.}{-240.}{-310.}
\caption{The distribution of: a) FIR luminosity; b) FIR luminosity of the 
cold component; c) FIR luminosity of the warm component; d) normalized FIR 
luminosity (to the K$^{\prime}$ band magnitude) for different Hubble types. 
The hatched histograms represent the distributions for the ``one component'' 
galaxies. The filled histograms represent the distributions for the galaxies 
(all BCDs) with detections only at two wavelengths (100 and 170\,${\mu}$m). 
From these, two BCDs don't have available K$^{\prime}$-band magnitudes and 
therefore they are excluded from the histogram (panel d, for Im-BCD).}
\end{figure}

\begin{figure}[htp]
\subfigure[]{
\includegraphics[scale=0.6]{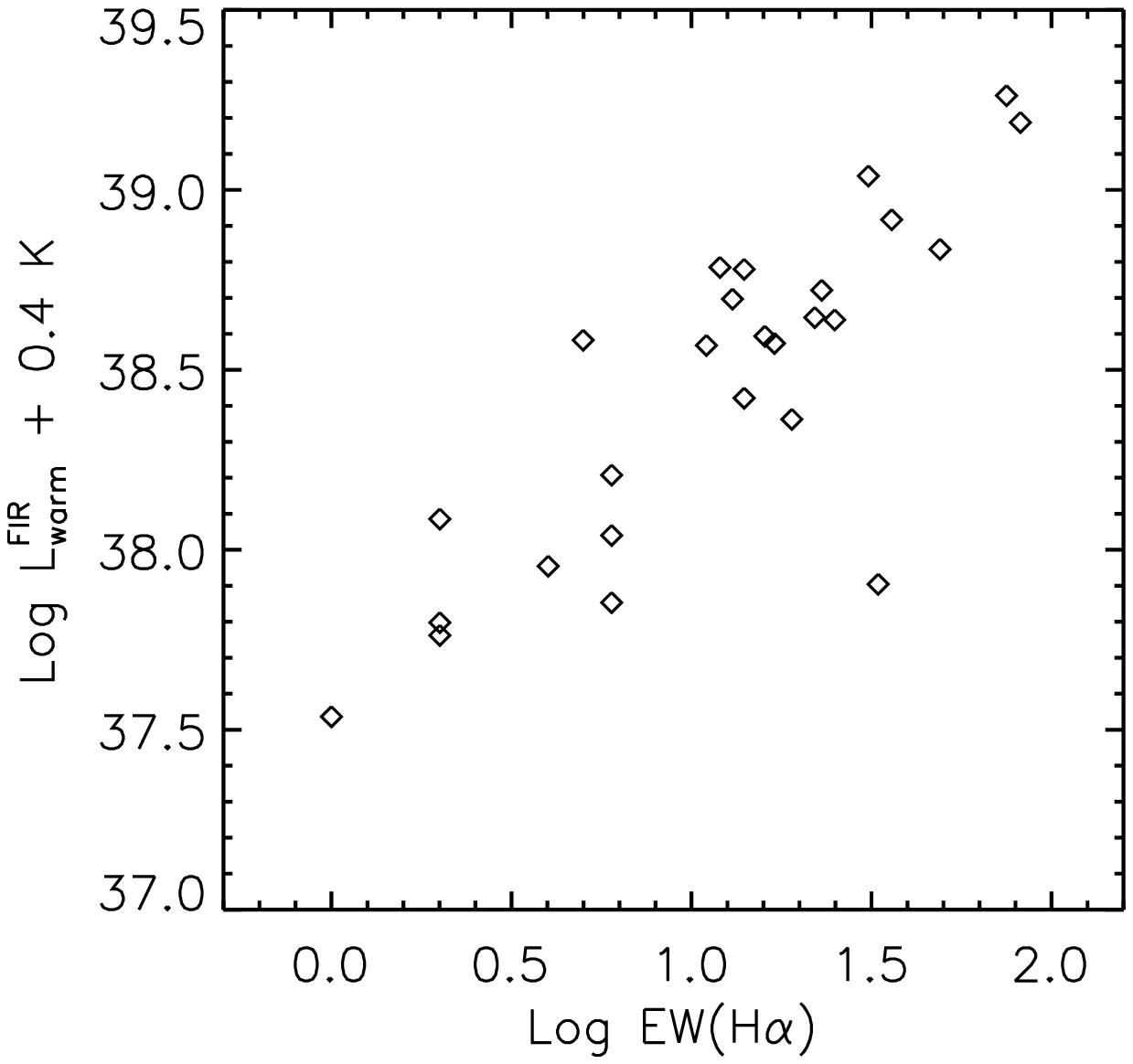}}
\subfigure[]{
\includegraphics[scale=0.6]{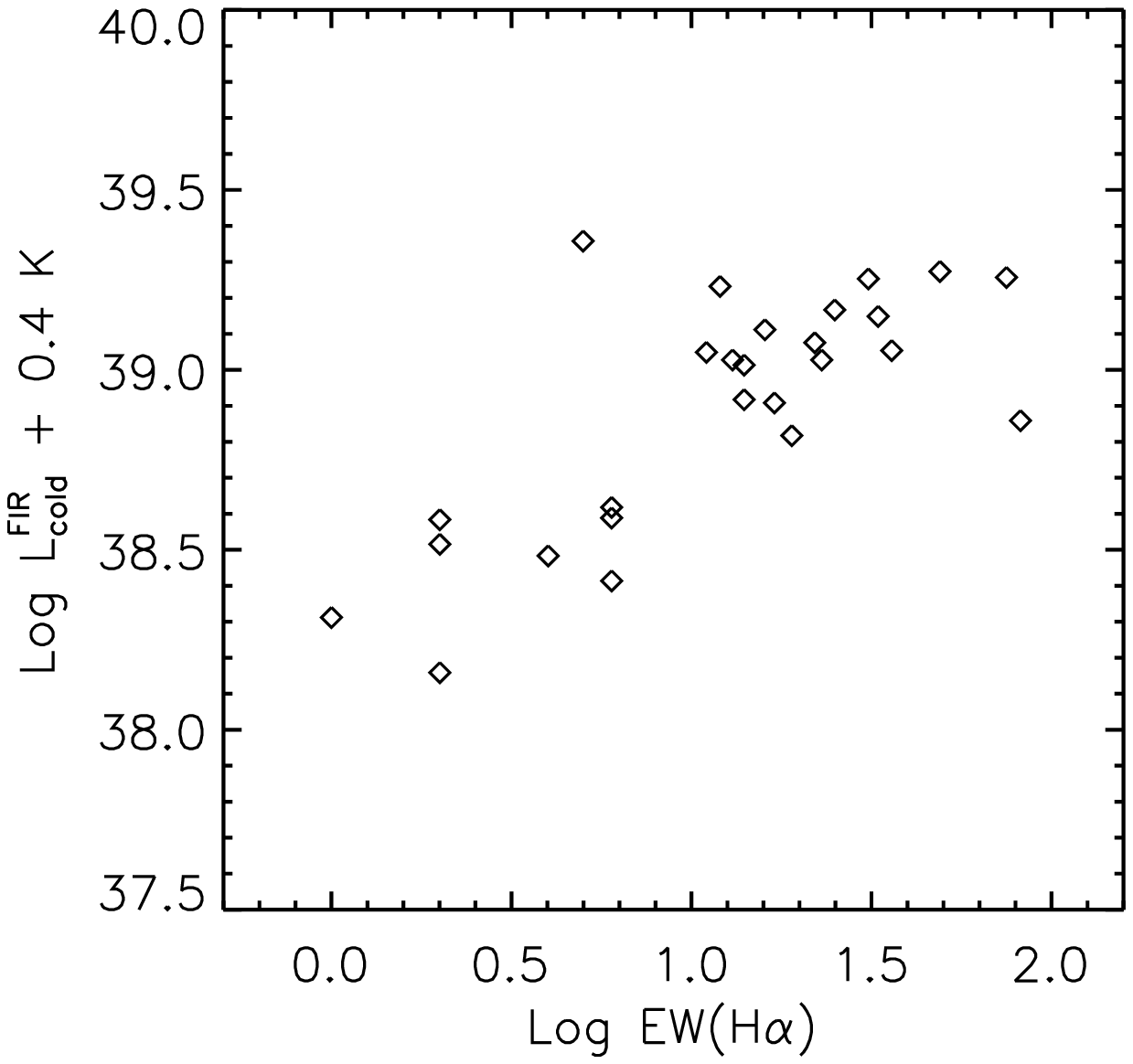}}
\caption{a) The normalized FIR luminosities of the warm dust component
(normalized to the K$^{\prime}$ band magnitudes) versus the H${\alpha}$ 
equivalent widths. 
b) The same for the cold dust component.}
\end{figure}

\begin{figure}[htp]
\subfigure[]{
\includegraphics[scale=0.6]{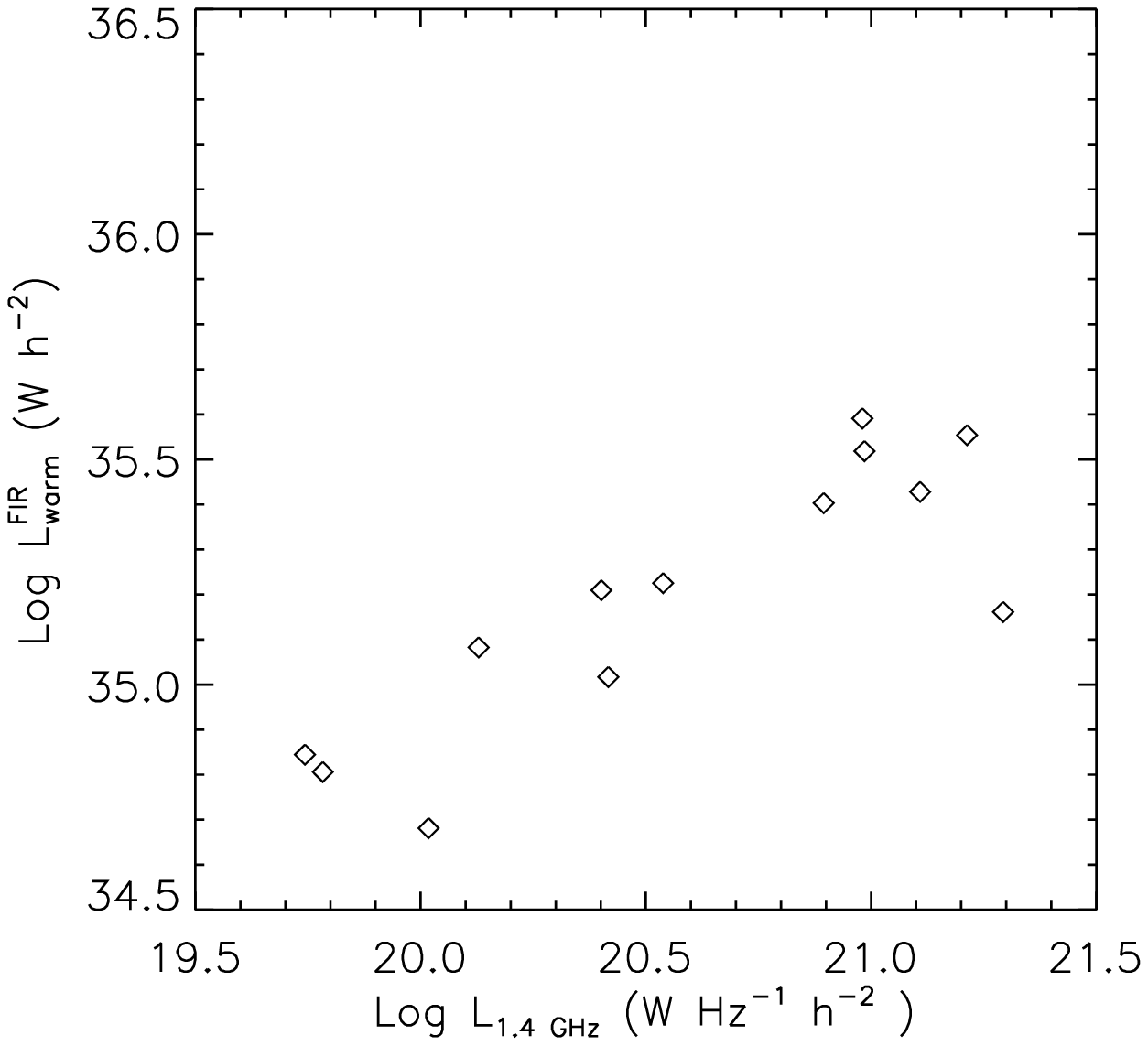}}
\subfigure[]{
\includegraphics[scale=0.6]{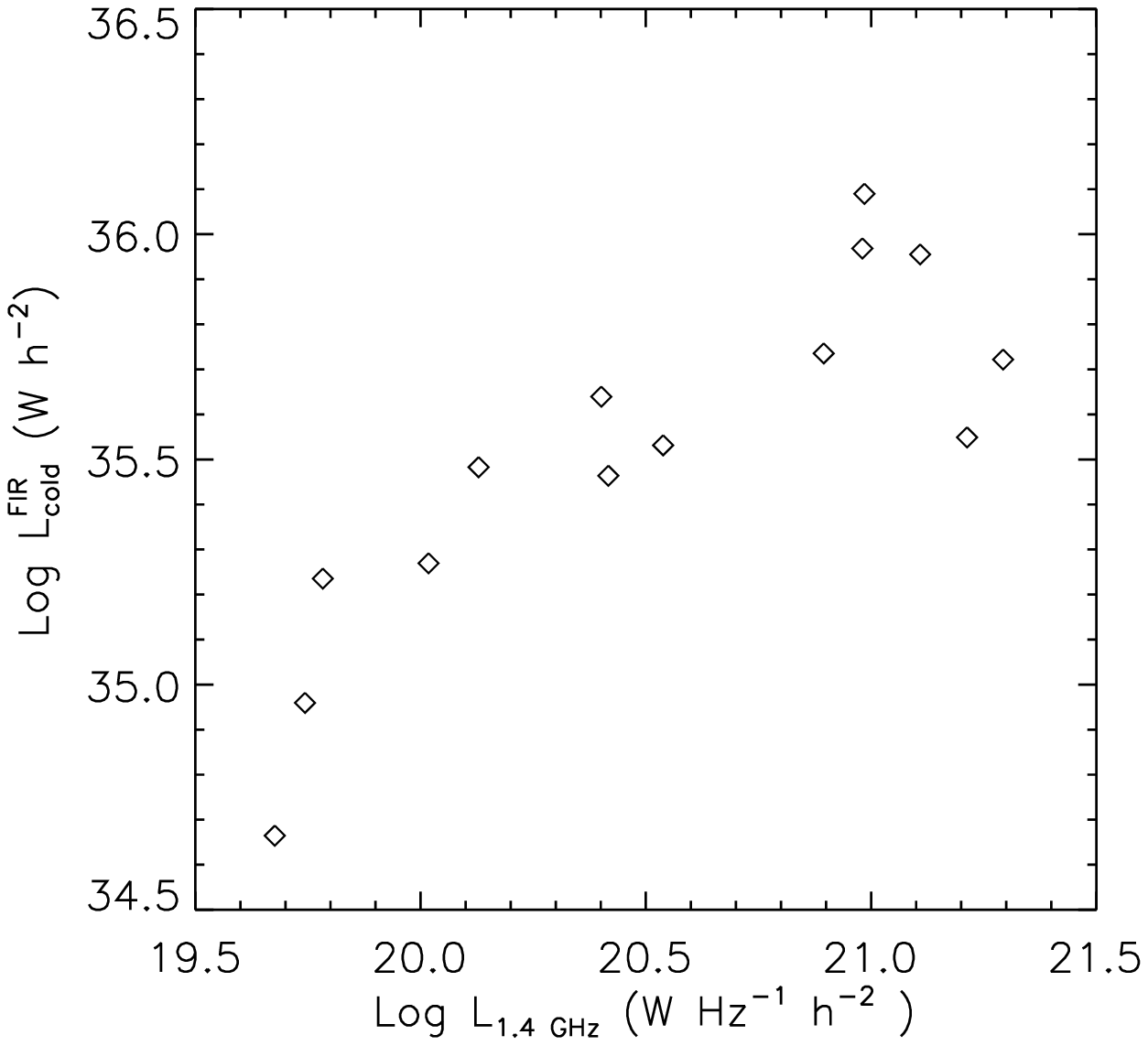}}
\caption{a) The warm FIR-radio correlation. 
b) The cold FIR-radio correlation.}
\end{figure}

\input{tab2}
\input{tab3}

\end{document}

%% file: eps.tex
\def\eps@scaling{.95}
\def\epsscale#1{\gdef\eps@scaling{#1}}

\def\plotone#1{\centering \leavevmode
    \epsfxsize=\eps@scaling\columnwidth \epsfbox{#1}}

\def\plotfiddle#1#2#3#4#5#6#7{\centering \leavevmode
    \vbox to#2{\rule{0pt}{#2}}
    \includegraphics{#1}}

%% file: tab2.tex
\begin{deluxetable}{ccccccccc}
\tablenum{2}
\tabletypesize{\small}
\tablewidth{480.pt}
\tablecaption{Results from constrained fits of two modified black-bodies to 
the 60, 100 and 170\,${\mu}$m flux densities}
\tablehead{
\colhead{VCC}                      &
\colhead{$M_{\rm D}^{\rm warm}$}           &
\colhead{$M_{\rm D}^{\rm cold}$}           &
\colhead{$T_{\rm D}^{\rm cold}$}           &
\colhead{$\sigma(T_{\rm D}^{\rm cold}$)\tablenotemark{(2)}} &
\colhead{$M_{\rm D}/M_{\rm HI}$\tablenotemark{(3)}}           &
\colhead{$L_{\rm FIR}$}                    &
\colhead{$L_{\rm FIR}^{\rm cold}$}         &
\colhead{note\tablenotemark{(4)}}                             \\
\colhead{}                                &
\colhead{M$_{\odot}\times {\rm h}^{-2}$\tablenotemark{(1)}}  &
\colhead{M$_{\odot}\times {\rm h}^{-2}$}  &
\colhead{K}                               &
\colhead{K}                               &
\colhead{}                                &
\colhead{W$\times{\rm h}^{-2}$}           &
\colhead{W$\times{\rm h}^{-2}$}           &
\colhead{}                                \\
\colhead{}                                & 
\colhead{$\times 10^{2}$}                 &
\colhead{$\times 10^{5}$}                 &      
\colhead{}                                &
\colhead{}                                &
\colhead{}                                &
\colhead{$\times 10^{34}$}                 &
\colhead{$\times 10^{34}$}                 &
\colhead{}                                \\
}
\startdata
   1 & 4.9 & 7.2 & 14.1 & 1.3 &\nodata & 1.6 & 0.95&\\ 
  66 & 120. & 140. & 15.5 & 0.4 & 0.0067 & 51. & 34.&\\ 
  87 & 4.9 & 8.9 & 14.0 & 1.0 & 0.0082 & 1.8 & 1.1&\\ 
  92 & 230. & 1100. & 13.6 & 0.2 & 0.0459 & 160. & 120.&\\ 
 152 & 73. & 43. & 18.5 & 0.5 & 0.0221 & 40. & 29.&\\ 
 318 & 6.1 & 16. & 14.9 & 0.5 & 0.0040 & 3.8 & 2.9&\\ 
 459 & 4.8 & 1.9 & 19.7 & 0.9 & 0.0024 & 2.5 & 1.9&\\ 
 460 & 110. & 38. & 20.2 & 0.9 & 0.0087 & 60. & 44.&\\ 
 655 & 20. & 260. & 12.6 & 0.2 & 0.6379 & 20. & 17.&\\ 
 664 & 38. & 10. & 15.7 & 2.5 & 0.0092 & 7.9 & 2.5&\\ 
 692 & 25. & 63. & 14.9 & 0.8 & 0.0513 & 16. & 12.&\\ 
 836 & \nodata   & \nodata   &\nodata&\nodata&\nodata& \nodata &  \nodata&see
Table 3\\ 
 857 & 28. & 110. & 15.4 & 0.3 & 0.0483 & 28. & 24.&\\ 
 873 & 180. & 190. & 16.1 & 0.2 & 0.0832 & 80. & 54.&\\ 
 912 & 28. & 8.9 & 20.5 & 1.2 & 0.0087 & 15. & 11.&\\ 
 971 & 12. & 8.5 & 17.8 & 0.5 & 0.0016 & 6.3 & 4.6&\\ 
1002 & 33. & 27. & 18.5 & 0.5 & 0.0110 & 23. & 19.&\\ 
1003 & \nodata & \nodata &\nodata&\nodata &\nodata&  \nodata&  \nodata&see
Table 3\\ 
1043 & 100. & 170. & 16.3 & 0.3 & 0.0906 & 67. & 53.&\\ 
1110 & 86. & 150. & 15.2 & 0.4 & 0.0711 & 43. & 30.&\\ 
1189 & 8.3 & 15. & 15.6 & 0.3 & 0.0103 & 4.8 & 3.6&\\ 
1253 & \nodata   & \nodata   &\nodata&\nodata&\nodata&  \nodata&  \nodata&see
Table 3\\ 
1326 & \nodata   & \nodata   &\nodata&\nodata&\nodata&  \nodata&  \nodata&see
Table 3\\ 
1379 & 45. & 26. & 18.4 & 0.5 & 0.0068 & 24. & 17.&\\ 
1410 & 10. & 8.1 & 15.7 & 0.7 & 0.0085 & 3.5 & 2.0&\\ 
1419 & \nodata   & \nodata   &\nodata&\nodata&\nodata&  \nodata&  \nodata&see
Table 3\\ 
1450 & \nodata   & \nodata   &\nodata&\nodata&\nodata&  \nodata&  \nodata&see
Table 3\\ 
1552 & 11. & 13. & 17.7 & 0.6 & 0.1446 & 8.1 & 6.5&\\ 
1554 & 250. & 45. & 19.0 & 0.5 & 0.0030 & 71. & 35.&\\ 
1575 & 49. & 13. & 18.5 & 1.3 & 0.0317 & 16. & 9.1&\\ 
1678 & 1.7 & 3.1 & 19.0 & 1.5 & 0.0006 & 2.8 & 2.5&\\
1686 & 18. & 6.1 & 19.3 & 0.8 & 0.0030 & 7.9 & 5.4&\\ 
1690 & 270. & 230. & 17.0 & 0.3 & 0.0745 & 130. & 93.&\\ 
1699 & 13. & 2.7 & 17.9 & 0.9 & 0.0012 & 3.3 & 1.5&\\ 
1725 & \nodata   & \nodata   &\nodata&\nodata&\nodata& \nodata &  \nodata&see
Table 3\\ 
1727 & 190. & 350. & 15.8 & 0.3 & 0.1144 & 120. & 90.&\\ 
1730 & 64. & 30. & 18.2 & 0.9 & 0.1134 & 28. & 18.&\\ 
1757 & \nodata   & \nodata   &\nodata&\nodata&\nodata&  \nodata&  \nodata&see
Table 3\\ 
\enddata
\tablenotetext{(1)}{${\rm h}={\rm H_0}/100$ and 
H$_0$ is the Hubble constant.}
\tablenotetext{(2)}{The uncertainty in the temperature of the cold dust
component.}
\tablenotetext{(3)}{The dust-to-HI-gas mass ratio.}
\tablenotetext{(4)}{In eight cases there was no evidence for two dust
components with different temperatures. For these galaxies there is a note that
sends the reader to Table~3, where these galaxies are listed.}
\end{deluxetable}

%% file: tab3.tex
\begin{deluxetable}{cccccc}
\tablenum{3}
\tabletypesize{\small}
\tablewidth{290.pt}
\tablecaption{Results from one modified black-body fit to the 60, 100 and
170\,${\mu}$m flux densities}
\tablehead{
\colhead{(1)} &
\colhead{(2)} &
\colhead{(3)} &
\colhead{(4)} &
\colhead{(5)} &
\colhead{(6)} \\
\colhead{VCC}                   &
\colhead{$M_{\rm D}$}           &
\colhead{$T_{\rm D}$}           &
\colhead{$\sigma(T_{\rm D})$} &
\colhead{$M_{\rm D}/M_{\rm HI}$}&
\colhead{$L_{\rm FIR}$}         \\
\colhead{}                         &
\colhead{M$_{\odot}\times {\rm h}^{-2}$}  &
\colhead{K}                               &
\colhead{K}                               &
\colhead{}                                &
\colhead{W$\times{\rm h}^{-2}$}           \\
\colhead{}                                &
\colhead{$\times 10^{4}$}                 &
\colhead{}                                &
\colhead{}                                &
\colhead{}                                &
\colhead{$\times 10^{34}$}                \\
}
\startdata
 130\tablenotemark{a} &  2.2  & 21.4 &  3.0& 0.0007 & 0.37\\ 
 836 &	120. &  27.1 &  0.3&0.0052 & 83.\\
 848\tablenotemark{a} &  10000.  &  9.0 &  0.5&  0.2026 & 9.1\\ 
1003 &	55. &  23.9 &  0.2&0.0061 & 17.\\
1253 &	36. &  20.6 &  0.8&0.0051 & 4.7\\
1326 &	7.2 &  34.7 &  1.2&0.0004 & 20.\\
1419 &	14. &  20.4 &  1.4&\nodata& 1.7\\
1450 &	92. &  22.1 &  0.3&0.0063 & 18.\\
1725 &	17. &  20.1 &  0.7&0.0026 & 1.9\\
1750\tablenotemark{a} &  5.1  & 17.8 &  1.9&  0.0050 & 0.27\\ 
1757 &	28. &  20.1 &  0.3&0.0282 & 3.0\\
\enddata
\tablenotetext{a}{VCC~130/848/1750 have detections only at 100 and
170\,${\mu}$m and therefore their dust masses represent lower limits and their
dust temperatures upper limits.}
\end{deluxetable}

%% file: ms.bbl
\begin{thebibliography}{}
\bibitem[]{} Alton, P. B., Bianchi, S., Rand, R. J., Xilouris, E. M., Davies,
J. I., \& Trewhella, M. 1998a, ApJ, 507, L125 
\bibitem[]{} Alton, P. B., Trewhella, M., Davies, J. I., Evans, R., 
Bianchi, S., Gear, W., Thronson, H., Valentijn, E., \& Witt, A. 1998b, 
A\&A, 335, 807
\bibitem[]{} Alton, P. B., Lequeux, J., Bianchi, S., Churches, D., Davies, J.,
\& Combes, F. 2001, A\&A, 366, 451
\bibitem[]{} Bianchi, S., Davies, J I., \& Alton, P.B. 2000a, A\&A, 359, 65
\bibitem[]{} Bianchi, S., Davies, J. I., Alton, P. B., Gerin, M., \& 
Casoli, F. 2000b, A\&A, 353, L13
\bibitem[]{} Binggeli, B., Popescu, C. C. \& Tammann, G. A. 1993, A\&AS, 
98, 275
\bibitem[]{} Binggeli, B., Sandage, A. \& Tammann, G. A. 1985, AJ, 90, 1681
\bibitem[]{} Boselli, A., Tuffs, R. J., Gavazzi, G., Hippelein, H., \& 
Pierini, D., 1997, A\&AS, 121, 507
\bibitem[]{} Bottinelli, L., Gouguenheim, L., Fouqu\`e, P., \& Paturel, G.  
1990, A\&AS, 82, 391
\bibitem[]{} Braine, J., Kr\"ugel, E., Sievers, A., \& Wielebinski, R. 1995, 
A\&A, 295, L55
\bibitem[]{} Braine, J., Gu\'elin, M., Dumke, M., Brouillet, N., Herpin, F.,
\& Wielebinski, R. 1997, A\&A, 326, 963
\bibitem[]{} Chini, R., Kreysa, E., Kr\"ugel, E., \& Mezger, P. G. 1986, 
A\&A, 166, L8 
\bibitem[]{} Contursi, A., Boselli, A., Gavazzi, G., Bertagna, E., Tuffs, R.,
\& Lequeux, J. 2001, A\&A, 365, 11
\bibitem[]{} de Jong, T., Klein, U., Wielebinski, R., \& Wunderlich, E. 1985, 
           A\&A, 147, L6
\bibitem[]{} Davies, J. I., Alton, P., Trewhella, M., Evans, R., \& 
Bianchi, S. 1999, MNRAS, 304, 495
\bibitem[]{} Devereux, N. A., \& Young, J. 1993, AJ, 106, 948
\bibitem[]{} Devriendt J. E. G., Guiderdoni B., \& Sadat R. 1999, A\&A, 350, 
381
\bibitem[]{} Draine, B. T. 1985, ApJS, 57, 587
\bibitem[]{} Draine, B. T., \& Lee, H. M. 1984, ApJ, 285, 89
\bibitem[]{} Dumke, M., Braine, J., Krause, M., Zylka, R., Wielebinski, R.,
\& Gu\'elin, M. 1997, A\&A, 325, 124
\bibitem[]{} Engargiola, G. 1991, ApJS, 76, 875
\bibitem[]{} Gavazzi, G., \& Boselli, A.,  1996, in: A UBVJHK photometric 
catalogue of 1022 galaxies in 8 nearby clusters, Gordon and Breach Science 
Publishers, New York
\bibitem[]{} Gavazzi, G., \& Boselli, A. 1999, A\&A, 343, 86
\bibitem[]{} Gu\'elin, M., Zylka, R., Mezger, P. G., Haslam, C. G. T., \& 
Kreysa, E. 1995, A\&A, 298, L29
\bibitem[]{} Gu\'elin, M., Zylka, R., Mezger, P. G., Haslam, C. G. T., 
Kreysa, E., Lemke, R., \& Sievers, A. W. 1993, A\&A, 279, L37
\bibitem[]{} Lisenfeld, U. \& V\"olk, H. J., 2000, A\&A, 354, 423
\bibitem[]{} Guiderdoni, B, \& Rocca-Volmerange, B. 1985, A\&A, 151, 108
\bibitem[]{} Haas, M., Lemke, D., Stickel, M., Hippelein, H., Kunkel, M.,
Herbstmeier, U., \& Mattila, K. 1998, A\&A, 338, L33
\bibitem[]{} Helou, G., Soifer, B. T., \& Rowan-Robinson, M. 1985, ApJ, 298, 
L7
\bibitem[]{} Helou, G., Khan, I. R., Malek, L., \& Boehmer, L. 1988, ApJS, 
68, 151
\bibitem[]{} Hoffman, G. L., Helou, G., Salpeter, E. E., Glosson, J., 
\& Sandage, A. 1987, ApJS, 63, 247
\bibitem[]{} Hoffman, G. L., Helou, G., Salpeter, E. E., \& Lewis, B. M. 
1989, ApJ, 339, 812
\bibitem[]{} Huchtmeier, W. K., \& Richter, O. G. 1986, A\&AS, 64, 111
\bibitem[]{} Hunter, D. A. \& Hoffman, L. 1999, AJ, 117, 2789
\bibitem[]{} Israel, F. P., Van Der Werf, P. P., \& Tilanus, R. P. J. 1999, 
A\&A, 344, L83
\bibitem[]{} Kennicutt, R. C., \& Kent, S. M. 1983, AJ, 88, 1094
\bibitem[]{} Kessler, M. F., Steinz, J. A.,
 Anderegg, M. E., Clavel, J.,
 Drechsel, G., Estaria, P., F\"alker, J.,
 Riedinger, J. R., Robson, A.,
 Taylor, B. G., \& Ximen\'{e}z de Ferr\'{a}n, S.
 1996, A\&A, 315, 27
\bibitem[]{} Kr\"ugel, E., Siebenmorgen, R., Zota, V., \& Chini, R. 1998, 
A\&A, 331, L9 
\bibitem[]{} Lemke, D., Klaas, U., Abolins, J.,
 \'{A}br\'{a}ham, P., Acosta-Pulido, J.,
 Bogun, S., Casta\~neda, H., Cornwall, L.,
 Drury, L., Gabriel, C., Garz\'{o}n, F.,
 Gem\"und, H. P., Gr\"ozinger, U.,
 Gr\"un, E., Haas, M., Hajduk, C.,
 Hall, G., Heinrichsen, I.,
 Herbstmeier, U., Hirth, G., Joseph, R.,
 Kinkel, U., Kirches, S., K\"ompe, C.,
 Kr\"atschmer, W., Kreysa, E.,
 Kr\"uger, H., Kunkel, M., Laureijs, R.,
 L\"utzow-Wentzky, P., Mattila, K.,
 M\"uller, T., Pacher, T., Pelz, G.,
 Popow, E., Rasmussen, I.,
 Rodr\'{\i}guez Espinosa, J., Richards, P.,
 Russell, S., Schnopper, H., Schubert, J.,
 Schulz, B., Telesco, C., Tilgner, C.,
 Tuffs, R. J., V\"olk, H. J., Walker, H.,
 Wells, M., \& Wolf, J. 1996,
 A\&A, 315, L64
\bibitem[]{} Lisenfeld, U., V\"olk, H. J., \& Xu, C. 1996, A\&A, 306, 677
\bibitem[]{} Mas-Hesse, J. M. \& Kunth, D. 1999, A\&A, 349, 765
\bibitem[]{} Matthews, T. A. \& Sandage, A. R. ApJ, 138, 30
\bibitem[]{} Melisse, J. P. M., \& Israel, F. P. 1994, A\&A, 285, 51
\bibitem[]{} Misiriotis A., Popescu, C. C., Tuffs, R. J., \& Kylafis, N. D. 
2000, A\&A, 372, 775  
\bibitem[]{} Neininger, N., Gu\'elin, M., Garc\'{\i}a-Burillo, S., Zylka, R.,
\& Wielebinski, R. 1996, A\&A, 310, 725 
\bibitem[]{} Popescu, C. C., Misiriotis A., Kylafis, N. D., Tuffs, R. J., 
\& Fischera, J., 2000, A\&A, 362, 138
\bibitem[]{} Preibisch, Th., Ossenkopf, V., Yorke, H. W. \& Henning, Th. 
1993, A\&A, 279, 577
\bibitem[]{} Siebenmorgen, R., Kr\"ugel, E., \& Chini, R. 1999, A\&A, 351, 
495
\bibitem[]{} Sievers, A. W., Reuter, H. -P., Haslam, C. G. T., Kreysa, E., 
\& Lemke, R. 1994, A\&A, 281, 691 
\bibitem[]{} Silva L., Granato G.L., Bressan A., \& Danese L. 1998, ApJ, 
509, 103
\bibitem[]{} Stickel, M., Lemke, D., Klaas, U., Beichman, C. A., 
Rowan-Robinson, M., Efstathiou, A., Bogun, S., Kessler, M. F., \& 
Richter, G. 2000, A\&A, 359, 865
\bibitem[]{} Trewhella, M., Davies, J. I., Alton, P. B., Bianchi, S., 
\& Madore, B. F. 2000, ApJ, 543, 153
\bibitem[]{} Tuffs, R. J. \& Gabriel, C. 2002, A\&A, in preparation
\bibitem[]{} Tuffs, R. J., Popescu, C. C., Pierini, D., V\"olk, H. J., 
Hippelein, H., Leech, K., Metcalfe, L., Heinrichsen, I., \& Xu, C. 2002, ApJS 
in press
\bibitem[]{} Tuffs, R. J., Lemke, D., Xu, C., Davies, J. I., Gabriel, C.,
Heinrichsen, I., Helou, G., Hippelein, H., Lu, N. Y., \& Skaley, D. 1996,  
A\&A, 315, L149
\bibitem[]{} Tully, R. B. \& Shaya, E. J. 1984, ApJ, 281, 31
\bibitem[]{} V\"olk, H. J. 1989, A\&A, 218, 67
\bibitem[]{} Warmels, R. H. 1988, A\&AS, 72, 427.
\bibitem[]{} Weaver, R., McCray, R., Castor, J., Shapiro, P., \& Moore, R. 
1977, ApJ, 218, 377
\bibitem[]{} Wunderlich, E., Wielebinski, R., \& Klein, U. 1987, A\&AS, 69, 
487
\bibitem[]{} Xilouris E. M., Kylafis N. D., Papamastorakis J., 
Paleologou E. V., \& Haerendel G. 1997, A\&A, 325, 135
\bibitem[]{} Xilouris E. M., Alton P. B., Davies J. I., Kylafis, N.,
Papamastorakis, J., \& Trewhella, M. 1998, A\&A, 331, 894
\bibitem[]{} Xilouris E. M., Byun Y. I., Kylafis N. D., Paleologou E. V., 
   \& Papamastorakis J. 1999, A\&A, 344, 868
\bibitem[]{} Xu, C. 1990, ApJ, 365, L47
\bibitem[]{} Xu C., \& Buat V. 1995, A\&A, 293, L65
\bibitem[]{} Xu C., \& Helou G. 1996, ApJ, 456, 163
\bibitem[]{} Xu C., Lisenfeld U., \& V\"olk H. J. 1994, A\&A, 285, 19
\bibitem[]{} Xu, C., Lisenfeld, U., V\"olk, H. J., \& Wunderlich, E. 1994, 
           A\&A, 282, 19
\end{thebibliography}
